\documentclass{iopjournal}
\usepackage{caption}
\usepackage[font=small]{caption}
\usepackage[english]{babel}
\usepackage{csquotes}
\usepackage{float}
\usepackage{booktabs}
\usepackage{indentfirst}
\usepackage{siunitx}
\usepackage{mathtools}
\usepackage{upgreek}
\usepackage{multirow}
\usepackage{rotating}
\usepackage{array}
\usepackage{ragged2e}
\usepackage{arydshln}  
\usepackage{enumitem} 
\usepackage{booktabs}
\usepackage{amsmath,amssymb}
\usepackage{graphicx}
\usepackage{subcaption} 
\usepackage[colorlinks=true, allcolors=blue]{hyperref}
\usepackage[symbol]{footmisc}

\begin{document}
\justifying
\articletype{Paper} 

\title{Demonstrating sub-picometer non-reciprocity levels in the Three-Backlink Experiment for LISA}
\author{Jiang Ji Ho-Zhang$^{1,2}$\footnote[1]{Author to whom any correspondence should be addressed.}\orcid{0000-0003-0379-8997}, 
Melanie Ast$^{1,2}$\orcid{0000-0003-1911-1686}, 
Lea Bischof$^{1,2}$\orcid{0000-0003-1576-7274}, 
Michael Born$^{1,2}$\orcid{0009-0001-5342-1779}, 
Daniel Jestrabek$^{1,2}$\orcid{0009-0004-9795-9388}, 
Stefan Ast$^{1,2,}$\footnote[2]{Present address: Deutsches Zentrum für Luft-und Raumfahrt e.V., Callinstraße 30b, 30167 Hannover, Germany}\orcid{0000-0003-4531-273X},  
Katharina-Sophie Isleif$^{1,2}$\footnote[3]{Present address: Helmut Schmidt University/University of the Federal Armed Forces Hamburg, Institute for Automation Technology, Holstenhofweg 85
22043 Hamburg, Germany}\orcid{0000-0001-7032-9440}, 
Oliver Gerberding$^{1,2}$\footnote[4]{Present address: Universität Hamburg, Institute for Experimental Physics, Luruper Chaussee 175, 22761
Hamburg, Germany}\orcid{0000-0001-7740-2698}, 
Thomas S. Schwarze$^{1,2}$\footnotemark[2]\orcid{0000-0002-2260-9971}, 
Jens Reiche$^{1,2}$\orcid{0009-0001-1147-9366},  
Gerhard Heinzel$^{1,2}$\orcid{0000-0003-1661-7868} 
and Karsten Danzmann$^{1,2}$\orcid{0000-0003-4174-683X}}

\affil{$^1$Max-Planck-Institut für Gravitationsphysik (Albert-Einstein-Institut), Callinstraße 38, 30167 Hannover, Germany}

\affil{$^2$Leibniz Universität Hannover, Welfengarten 1, 30167 Hannover, Germany}

\email{\hyperlink{mailto}{jiangji.hozhang@aei.mpg.de}}


\begin{abstract}
\justifying
{\fontsize{10pt}{12pt}\selectfont 

The current planned space-based gravitational-wave detectors require a bidirectional optical connection, referred to as Backlink, between two adjacent optical benches to provide a mutual phase reference for the local interferometric measurements. However, if the Backlink shows asymmetry between the two propagation directions, the effective optical pathlengths of the counter-propagating beams can introduce a differential phase noise, called non-reciprocity, into the main interferometric measurement that will limit the achievable accuracy in time-delay interferometry (TDI) post-processing. Hence, it is important to understand the properties of the Backlink to ensure that it will not compromise the interferometric detection. The Three-Backlink Experiment (3BL), which consists of an optical test facility with two rotatable benches, was designed under the Laser Interferometer Space Antenna (LISA) framework to study the performance of three Backlink configurations: two fiber-based and one free-beam scheme. In this paper, we report recent experimental results from the 3BL. We describe the commissioning and the subsequent noise mitigation. We achieve a setup noise floor below $1  \text{ pm}/\sqrt{\text{Hz}}$ across most of the LISA measurement band, and provide an understanding of the current technical limitations. With this low-noise baseline, we measured the performance of the three Backlink implementations under non-rotational conditions. We show that all three Backlinks reach sub-picometer non-reciprocity levels across most of the frequency band, with the remaining part dominated by the mentioned testbed noise. This enabled us to conduct a preliminary study of the Backlink inherent noise, where we emphasized on the backscatter noise intrinsic to a straightforward fiber-based Backlink, as this is the current baseline for LISA.}
\end{abstract}

\section{Introduction}
\smallskip

The  Laser Interferometer Space Antenna (LISA) is the first planned space-based gravitational wave detector \cite{LISA:2024hlh}. The mission is led by the European Space Agency (ESA) and was formally adopted in January 2024. It aims to detect signals in the low frequency band, between 0.1 mHz and 1 Hz, using highly sensitive laser interferometry at 1064 nm. LISA will consist of three spacecraft arranged in a quasi-equilateral triangular formation with arm lengths of 2.5 million kilometers. Each satellite will send two laser beams to the respective distant satellites, creating a total of six laser links. To maintain the triangular formation despite the relative displacement induced by the orbital dynamics ---which result in angular variations on the order of $1.5^\circ$ per year---, each satellite is equipped with two moving optical sub-assemblies (MOSAs) with one optical bench per unit. These optical benches are interconnected by an optical link known as the Backlink (BL), or Phase Reference Distribution System (PRDS), which provides a mutual phase reference between them for the local interferometric measurements. Hence, two counter-propagating beams travel through the Backlink. Ideally, the forward and backward beams see the same effective optical pathlengths, resulting in zero differential phase. However,  in reality, there will be residual noise between the two directions of propagation, characterized by a single parameter defined as the non-reciprocity of the Backlink. Any difference in the measured pathlengths will contribute as a differential phase noise, or non-reciprocal phase noise, that will propagate into the LISA interferometry. To minimize this effect, the non-reciprocity must be kept very low. Specifically, it must be below 1 \text{pm}/$\sqrt{\text{Hz}} \cdot u_\text{}(f)$ or, equivalently, $ 6~\upmu \text{rad}/\sqrt{\text{Hz}} \cdot u_\text{}(f)$,  within the LISA measurement band, where $u_\text{}(f) = \sqrt{1+(2.0\,\text{mHz}/f)^4}$ is the noise shape function defined by the mission requirements and $f$ is the Fourier frequency.

The feasibility of such a low-noise optical link has been investigated in previous experiments. The results of the first single-bench, fiber-based Backlink experiment \cite{FleddermannPhD}, together with the experimental investigations of the fiber dynamics \cite{RohrPhD, paperMax2020}, provide initial evidence that a straightforward connection is capable to meet the requirements. In addition, proof‑of‑concept studies on a free-beam Backlink connection \cite{IsleifPhD, Chilton:2019qdb} demonstrated the viability of this design. Several follow-up Backlink designs have been thoroughly investigated, as presented in \cite{IsleifPhD}. The three most promising candidates were selected and integrated into one single setup called the Three-Backlink Experiment (3BL), which enables simultaneous optical testing of the three Backlink concepts. Two of them, referred to as the Direct Fiber Backlink (DFBL) and the Frequency-Separated Fiber Backlink (FSFBL) are fiber-based, while the third propagates in free space and is referred to as the Free-Beam Backlink (FBBL). Understanding the limitations inherent to each design is essential for the Backlink integration in any space-based gravitational-wave detector. In particular, the missions Tianqin \cite{TianQin:JLuo}, Taiji \cite{Taiji:ZLuo}, and DECIGO \cite{Decigo:Kawamura}, as well as other future LISA-like missions, can strongly benefit from the results achieved with the 3BL, as the specifications might vary and, consequently, differ on the Backlink implementation requirement. For LISA, a successful demonstration of the DFBL performance offers additional experimental validation to support the current baseline implementation of a direct fiber-based Backlink.

In this paper, we report the current performance of the Three-Backlink Experiment. Our results indicate that all three Backlink implementations are capable of reaching the demanding performance level of 1 \text{pm}/$\sqrt{\text{Hz}} \cdot u_\text{}(f)$. We present a characterization of the fundamental and technical noise sources of the experimental setup, and the corresponding reduction and mitigation strategies. Additionally, we provide a preliminary study of the non-reciprocity performance of each Backlink design. Finally, we conclude with an analysis of the relevance of the DFBL performance results to the LISA mission.

\section{The Three-Backlink Experiment}
\smallskip

\begin{figure}[h] 
    \centering
        \includegraphics[width=0.9\linewidth, trim= 20 10 10 30, clip]{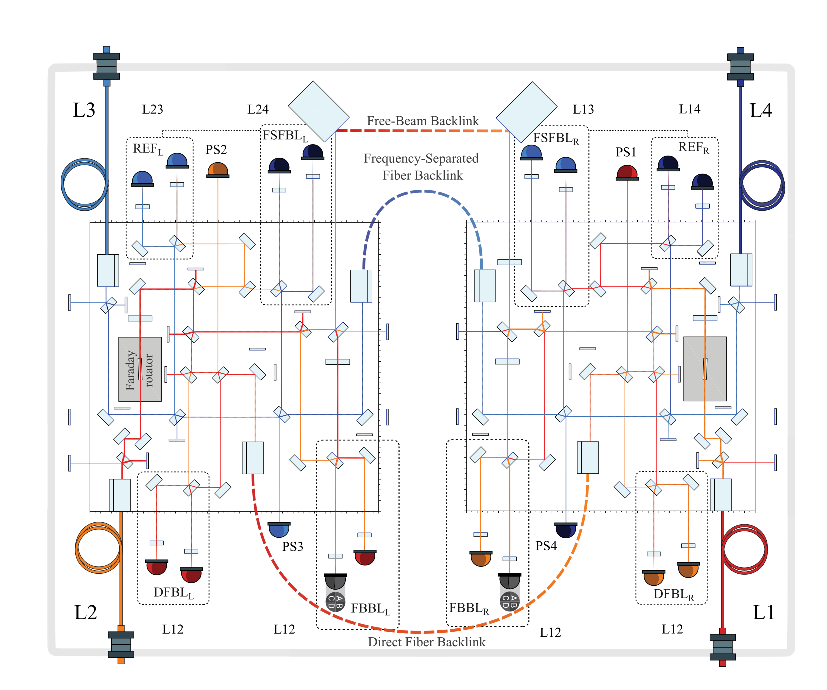}
    \caption[Schematic of the 3BL optical design]{Schematic of the 3BL optical design. The three Backlink connections are illustrated as dashed lines, with the respective interferometers indicated within a dashed box (DFBL, FSFBL+REF, FBBL) on the left (L) or right (R) bench. The four lasers involved in the experiment are defined as L2 and L3 in the left bench, and L1 and L4 in the right bench. Above each interferometer, the interfering lasers are shown, resulting in the corresponding beatnotes, e.g. L12. All four lasers are frequency stabilized to a fifth laser in a separate setup, not shown here. The colored lines show the optical laser beam paths. Additionally, the in-loop photodetectors for the power stabilization (PS) are represented. The figure is an adaptation from \cite{IsleifPhD}.}
    \label{fig:3BLoptocad}
\end{figure}

The Three-Backlink Experiment is an optical testbed designed to evaluate the performance of three Backlink implementations for LISA. It consists of two quasi-monolithic optical benches that are rotatable by $\pm 1.5^\circ$, which enables the simulation of the angular breathing in LISA. The Backlink candidates, selected from a dedicated trade-off analysis that considered the architecture complexity and the straylight contribution \cite{IsleifPhD}, are the Direct Fiber Backlink, the Frequency-Separated Fiber Backlink and the Free-Beam Backlink. A detailed description of each is provided in the accompanying paper to this work \cite{paperLea25} and is therefore omitted here. The design of the 3BL optical layout is shown in figure \ref{fig:3BLoptocad}. The three Backlink connections are represented by dashed lines and labeled accordingly, with the respective interferometers indicated within a dashed box (DFBL, FSFBL+REF, FBBL) on the left (L) or right (R) bench. The four lasers involved in the experiment are defined as L2 and L3 in the left bench, and L1 and L4 in the right bench, and are locked to the same (fifth) reference laser, all at different frequencies with kilohertz offsets between them. The beatnote combinations in the setup are labeled accordingly next to each interferometer (e.g., DFBL receives light from L1 and L2, generating the beatnote at frequency L12). For clarification, as noted in \cite{paperLea25}, L3 and L4 are employed in the FSFBL design to achieve a frequency separation to mitigate the impact of Backlink fiber backscattered light, necessitating two additional interferometers, REF, to cancel the phase noise of the additional lasers.

The construction of the 3BL was conducted in-house, at the Albert Einstein Institute, under cleanroom conditions. It delivered two quasi-monolithic Clearceram baseplates with fused silica optical components bonded with UV-glue, implementing the also in-house developed quasi-monolithic fiber collimators, or fiber injector optical subassembly (FIOS). A comprehensive description of the process and analysis of the final alignment results is provided in \cite{LeaBischofPhD, KnustMaster2020,paperLea25}. The experiment is conducted under vacuum conditions, while all the electronics and a dedicated laser preparation setup are placed outside of the vacuum chamber. Further details of the laboratory infrastructure can be found in \cite{IsleifPhD, LeaBischofPhD,paperLea25}. The modifications with respect to the cited works were implemented to meet the current performance, which will be presented in section \ref{sec:commissioning}. 

Due to the two separated benches in the 3BL, it is not possible to have a stable, quasi-monolithic reference interferometer on the bench to cancel out the relative phase noise of L1 and L2 (while for L3 and L4 it is canceled with the REF interferometers). Hence, this will remain in the individual Backlink non-reciprocal phase noise measurement. Assuming that the two benches are fully symmetric, this will be: 
\begin{equation}
\label{eq:singleBL}
    \begin{aligned}
   {\Phi_{\text{BL}}}:={\varphi^{\text{L}}_{\text{BL}}} + {\varphi^{\text{R}}_{\text{BL}}}   =  {\varphi_{\text{NR,BL}}} + 2[{\varphi_{\text{L2}}}-{\varphi_{\text{L1}}}]  ,\\ 
\end{aligned}
\end{equation}
where ${\varphi^{\text{L}}_{\text{BL}}}$  and ${\varphi^{\text{R}}_{\text{BL}}}$  are the interferometric phases measured on the left and right bench, respectively,  ${\varphi_{\text{NR,BL}}}$ is the non-reciprocal phase noise of the Backlink, and  ${\varphi_{\text{L1}}}$  and  ${\varphi_{\text{L2}}}$  are the phase noises of the L1 and L2 lasers. The combination includes the term  $2[{\varphi_{\text{L2}}}-{\varphi_{\text{L1}}}]$, which will dominate and mask the actual non-reciprocity measurement. This apparent limitation is resolved when the differential non-reciprocities between two Backlinks are calculated, resulting in:
 \begin{equation}
     \label{eq:nonRec}
        \begin{aligned}
{\Phi_{\text{BL1}}  - \Phi_{\text{BL2}}} = {\varphi_{\text{NR,BL1}} - {\varphi_{\text{NR,BL2}}}} .\\ 
   \end{aligned}
 \end{equation}

Details on the derivation of equations (\ref{eq:singleBL}) and (\ref{eq:nonRec}) are found in \cite{IsleifPhD}. Hence, the measurements obtained with the 3BL represent the differential non-reciprocities between two Backlinks, forming three pairs of possible combinations. With these combinations, it is possible to disentangle the dominant noise sources of each Backlink in the respective combination.

\section{Commissioning}
\label{sec:commissioning}
\smallskip

To evaluate the performance of each Backlink design, it is essential to first identify the fundamental and technical noise limitations of the environment and experimental setup. The initial commissioning phase and its results are presented in \cite{paperLea25}, which showed a performance below 15~\text{pm}/$\sqrt{\text{Hz}} \cdot u_\text{}(f)$ over the entire frequency band, establishing a solid baseline for further investigations. In this section, we describe the follow‑up commissioning phase, presenting the characterized noise sources together with their corresponding mitigation strategies. The contribution of each noise to the final noise budget will be presented in section \ref{sec:results}.

\subsection{Readout Noise}
\label{sec:readoutNoise}
\smallskip
The readout noise comprises contributions from the transimpedance amplifier and the readout Phasemeter. 

\subsubsection{Phasemeter}
\label{sec:pm_hfnoise}

Two Phasemeters are implemented in the 3BL, namely the locking Phasemeter and the readout Phasemeter. The former is used to frequency stabilize the four lasers to the reference and has been already characterized in \cite{LeaBischofPhD}. Hence, we focus only on the readout Phasemeter, hereafter referred to simply as the Phasemeter, which measures the interferometric phases of all three Backlinks. 

The Phasemeter extracts the amplitude and phase from the respective heterodyne signals. It consists of an IQ-demodulation extended with a phase-locked loop (PLL), which enables the Phasemeter to track the signals in larger dynamic ranges \cite{GeberdingPhD}. In the 3BL, the Phasemeter is optimized for kilohertz frequencies and it consists of sixteen Analog-to-Digital Converters (ADCs) with an input range of $2\,\mathrm{V_{pp}}$, 20 bits of resolution, and a sampling rate of 1 MHz \cite{RischkopfMaster2021}. One particular noise that couples in the PLL is the so-called additive noise \cite{SchwarzePhD, BodePhD}:
\begin{align} \tilde{\varphi}_{\text{add}}(f) &=  \frac{\tilde{a}_{\text{in}}(f)}{A_{\text{RMS}}}, \end{align}
where  $\tilde{\varphi}_{\mathrm{add}}(f)$ is the resulting phase noise, ${\tilde{a}_{\mathrm{in}}}(f)$ is the amplitude noise around the carrier signal frequency, ${A_{\mathrm{RMS}}}$ is the RMS signal amplitude, and the tilde denotes that the variable is expressed as an amplitude spectral density (ASD). As can be seen, the additive noise is inversely proportional to the amplitude of the input signal. Hence, enhancing the amplitude within the ADC range reduces the white noise accordingly. In addition to this, our measurements also indicate that the Phasemeter dark noise is reduced when the amplitudes of the input signals are closely matched. We hypothesize that noise suppression is more effective under these conditions, but this requires further validation. An individual adjustment of each amplitude is possible with the current design of the transimpedance amplifier.

\subsubsection{Transimpedance Amplifier}
\label{sec:TIA_HF}

The transimpedance amplifier (TIA) converts the photocurrent of each heterodyne signal into a voltage signal, which is later sent to the Phasemeter. By adjusting the feedback gain of each channel, the amplitudes can be increased and matched as needed. However, the current TIA design exhibits parasitic capacitance, which introduces a low-pass filter behavior with a cut-off frequency at around  $500\,\text{kHz}$. Several attempts have been made to modify the electronic circuit design but the effect of the parasitic capacitance persists. Consequently, all interferometric signals are restricted to frequencies below this threshold, as higher frequency values result in attenuation of the signal and substantial increase in white noise. 

A further noise reduction has been achieved by operating each Backlink in a separate TIA housing. Previously, only two in-house built 16-channel TIA devices were used, as they would cover all 22 photodiode channels, including single and quadrant photodiodes. However, this led to cross-talk between signals with frequencies near fundamental or harmonic values relative to another. Hence, the electrical separation of the signals at this stage reduced effectively the overall noise. 

Finally, the design of the TIA electronic circuit has been modified to provide better isolation between channels. Previously, high noise levels present in one signal would couple into the others. Now, the improved decoupling ensures clearer signal separation.

\subsubsection{Combined Noise}
\label{sec:readoutNoise_LF}
Here, we evaluate the combined contribution of the TIA and the Phasemeter to the readout noise. In \cite{LeaBischofPhD}, a lower limit in the optimal frequency range of the Phasemeter performance was observed to be at around $200\,\text{kHz}$, with the underlying cause still under investigation. With the upper limit of 500 kHz imposed by the TIA, this establishes an effective range of frequencies between $200\,\text{kHz}$ and $500\,\text{kHz}$ for the five beatnote frequencies involved in the experiment. However, this creates a performance limitation at frequencies below $\mathit{f} < 10 ^{-2}\  \mathrm{Hz}$. To quantify this effect, optical-electrical split tests have been performed. Optical-electrical split test refers to a measurement where the photocurrent of one measurement signal is split into $n$ equal photocurrents and sent to $n$ TIA channels with equal gains. Hence, the Phasemeter receives ideally $n$ identical signals originated from the same optical source. This way, the split test measurement include both TIA and Phasemeter noises, which should cancel out in the subtraction. The measurements confirmed the expected low-frequency noise limitation. For increasing signal frequencies, the readout noise increases proportionally. At frequencies close to 500~kHz, the readout noise reaches the 1 \text{pm}/$\sqrt{\text{Hz}} \cdot u_\text{}(f)$ requirement. This sets a clear limitation particularly for FSFBL, as its design requires a larger number of beatnotes at separated frequencies, including some that are closer to the upper limit. 

\subsection{Detection Noise} 
\label{sec:detectionNoise}
\smallskip
The detection noise comprises the contributions introduced at the photodetectors. 

\subsubsection{Additive detection noise}

We define and characterize the contributions to the phase noise of the relative intensity noise (RIN), the shot noise (SN), and the electronic noise (EN), which are expected to introduce a white noise contribution across the entire frequency range. Assuming uncorrelated RIN between the two interfering beams, the induced phase noise is \cite{paperLennart2022}: 
\begin{align} {\varphi}_{\text{RIN}} &=   \sqrt{\frac{P^2_{\text{meas}} +{P^2_{\text{ref}}}}{2 \eta_{\text{het}} P_\text{meas} P_\text{ref}}}  \ \tilde{r}(1f_{het}), \end{align}
while the phase noises induced by SN and EN are derived from \cite{Tröbs2012}:
\begin{align}
{\varphi}_{\text{SN}} &= \sqrt{ \frac{q_e \left(P_\text{meas} + P_\text{ref} \right)}{R \eta_{\text{het}} P_\text{meas} P_\text{ref}} }, \\
{\varphi}_{\text{EN}} &= \frac{2n_\text{el}}{R \sqrt{\eta_{\text{het}} P_\text{meas} P_\text{ref}} } ,
\end{align}
where $P_\text{meas}$ and $P_\text{ref}$ are the measurement and reference beam powers, $\eta_{\text{het}}$ is the heterodyne efficiency, \mbox{$R \approx 0.7$ A/W} is the photodiode responsitivity, $q_e = 1.6\cdot10^{-19}$ A$\cdot$s  is the elementary charge, and $n_{\text{el}}= 5\cdot10^{-12} $A$\cdot\mathrm{Hz}^{-1/2}$ is the equivalent current noise ASD of the photodiode. We measured the  $1f$-RIN ASD, $\tilde{r}(1f_{\text{het}})$, for all lasers and adopted the worst-case value of $1 \cdot 10^{-7}\,\text{Hz}^{-1/2}$ in the calculations. Additionally, the contribution of $2f$-RIN \cite{paperLennart2022} resulted to be negligible and is thereby not considered. Table \ref{RN_SN_EN_table} shows the resulting phase noise induced by RIN, SN, and EN, for each interferometer. For simplicity, only one photodiode per interferometer is shown, as the second differs minimally in power and yields very similar values. In section \ref{sec:results}, these results will be shown as a single scalar, as they introduce white noise, across the entire frequency range, corresponding to the uncorrelated sum of the individual contributions present in each Backlink combination. 

\begin{table}[htbp]
\centering
\begin{tabular}{cccc}
\toprule
& \multicolumn{3}{c}{Phase noise  $(\upmu\text{rad}\,\text{Hz}^{-1/2})$} \\
Interferometer &  ${\varphi}_{\text{RIN}}$  & $ {\varphi}_{\text{SN}} $ & $ {\varphi}_{\text{EN}}$ \\\midrule
$\mathrm{DFBL_{L}}$ & $0.322$ & $0.181$ & $0.404$ \\
$\mathrm{DFBL_{R}}$ & $0.312$ & $0.166$ & $0.356$ \\
$\mathrm{FBBL_{L}}$ & $0.151$ & $0.081$ & $0.137$ \\
$\mathrm{FBBL_{R}}$ & $0.131$ & $0.070$ & $0.117$ \\
$\mathrm{FSFBL_{L}}$ & $0.187$ & $0.074$ & $0.113$ \\
$\mathrm{FSFBL_{R}}$ & $0.352$ & $0.135$ & $0.205$  \\
$\mathrm{REF_{L}}$ & $0.113$ & $0.041$ & $0.047$  \\
$\mathrm{REF_{R}}$ & $0.142$ & $0.053$ & $0.063$ \\
\bottomrule
\end{tabular}
\caption{Calculated phase noise contributions due to RIN, SN and EN. Subindex L and R denote that the interferometer is on the left or right bench, respectively.}
\label{RN_SN_EN_table}
\end{table}
\subsubsection{Detector Crosstalk}
\label{sec:detectorCrosstalk}
Strong evidence of crosstalk has been identified between the photodetectors. Initially, they shared a common bias voltage distributed across three connector structures inside the vacuum chamber. Due to unshielded cables and a lack of insulation within the bias distributor, crosstalk between the photodiode signals occurred at this point. To address this issue, the photodiode design has been updated to incorporate decoupling capacitors in the bias connectors. Additionally, shielded cables are used and an enhanced bias distributor has been designed, which ensures the isolation of every signal.

\subsection{Straylight}
\label{sec:straylight}
\smallskip
Straylight mitigation is required in any optical experiment to avoid undesirable interferences at the same frequencies as the measurement signals. In the 3BL, a thorough investigation on straylight mitigation at the optical benches was conducted in \cite{IsleifPhD}, which eliminated most straylight beams by design. In this section, we focus on the further reduction of straylight present in the experimental setup, occurring at the laser preparation table and inside the thermal shield, which is the enclosure for the optical benches (see \cite{IsleifPhD, paperLea25} for detailed description of the experimental setup). For both cases, a dedicated infrared laser visualizer was used to reveal the straylight. For the purpose of clarification, it is important to not confuse the straylight described in this section with the intrinsic straylight of the Backlink fiber, which has a dedicated analysis in section \ref{sec:fiberbasednoise}.

At the laser preparation stage, all lasers are set up on a shared optical table, where the necessary adjustments are performed before they enter the vacuum chamber. We observed that a fraction of the light was leaked through the cladding of the fibers right after the fiber couplers and at the inputs of the variable optical attenuators (VOAs), producing a uniform glow on the fiber. Due to the setup configuration, this lead to diffused scatter influencing the neighboring optical layouts. To mitigate this effect, we implemented black walls in the setup that isolate the optical path of each laser. 

Inside the in-vacuum setup, which is surrounded by a thermal shield, we observed that the photodetector surfaces generated secondary reflections. While the photodiode mounts are purposely angled by $8^\circ$ to prevent a direct coupling of these reflections, some fraction can be redirected and coupled back into the nominal beam paths via the reflecting surfaces of the shield. To mitigate this effect, we covered these surfaces with a vacuum-compatible optical black foil, designed specifically to suppress straylight. 

Despite these efforts, our measurements indicate the presence of residual straylight, evidenced by the characteristic shoulder in the noise spectrum in two Backlink implementations, the DFBL and the FBBL. This contribution is canceled with balanced detection (see section \ref{sec:results}), suggesting that it is not originated by diffused straylight. Hence, a potential explanation is that a beatnote is formed at an early stage, interfering with the nominal beam and propagating through the experiment. Our current assumption attributes this effect to a parasitic cross-bench coupling between the L1 and L2 beams, likely occurring through the high-power beams at the first optics due to reflections on the surfaces of the Faraday rotator or the thermal shield. To mitigate this possible effect, a separation between the two optical benches with additional optical black foil is planned for future work. 

\subsection{Environmental Temperature}
\label{sec:EnvTemp}
\smallskip

Initial studies of the temperature coupling in the 3BL experiment are reported in \cite{LeaBischofPhD}. In this section, we present the follow-up investigations. During the preliminary study, a scroll pump was used to reach vacuum conditions of $10^{-4}$ mbar for the experiment. However, this induced thermal gradients in the laboratory, despite environmental control by the Heating, Ventilation, and Air Conditioning (HVAC) system, that required a long time to propagate to the in-vacuum setup. Consequently, this resulted in prolonged ongoing temperature drifts at the Backlink fiber that lead to up-conversion of this low-frequency drift into the LISA band, visibly increasing the noise towards lower frequencies (see section \ref{sec:T2Pscatter}). Hence, we replaced the scroll pump by the central vacuum system of the building, mitigating completely the effect of the temperature gradient. Additionally, to reduce external temperature coupling into the vacuum chamber, we replaced its original insulating material, Styrodur, with Armaflex. Although both feature similar thermal conductivity --which would effectively not change the final performance--, the advantage of Armaflex is that it offers double the thickness, a strong adhesion, and more flexibility, ensuring full coverage of the tank surface and sealing all potential gaps to prevent air flow. Styrodur is rigid, potentially leading to inadequate coverage and creating gaps in the insulation. In-vacuum temperature sensors measure a temperature stability of $\mathrm{\tilde{s}^{3BL}_\text{TF}=10{\ \upmu K/\sqrt{Hz}} \cdot \mathit{u_\text{}(f)}}$, which is better than the required for the LISA Backlink fiber. With this level of stability and a low temperature drift, the experiment operates under specification. 

Finally, we observe a potential coupling of ambient temperature fluctuations into our measurements. Although a direct confirmation has not yet been obtained, this is corroborated by strong correlations between the data of the HVAC and the non-reciprocity measurements, as the time series exhibit the same periodic oscillation, matching with the temperature‑control cycle. While the exact coupling mechanism remains under investigation, our current hypothesis places the coupling outside the vacuum chamber, most likely in the Phasemeter, as this is the common element across all measurements showing this effect. The in-vacuum temperature sensors also show this oscillation, but the effect is below the requirement. The impact on the performance is visible at low frequencies (see section \ref{sec:results}) and further mitigation strategies are planned for the near term. 

\subsection{Laser Noise}
\label{sec:LaserNoise}
\smallskip

To minimize contributions of laser noise to the measurements, the power and frequency of all lasers are stabilized to meet the performance requirements. Here, we quantify the impact of each contribution by projecting the measured power and frequency noise into the measured non-reciprocity. 

\subsubsection{Power Noise}
Laser power fluctuations, particularly at the DC level, couple directly into the phase measurements. In the 3BL, the exact coupling mechanism is still under investigation. To quantify this effect, we determined the power-to-phase coupling coefficients of every laser in the respective Backlink combinations. For this purpose, a well-known modulation from a standard signal generator is added to the power of the laser through a variable optical attenuator (VOA), which is implemented in the setup purposely to adjust the powers in the experiment. The modulation has been applied at various frequencies ($\mathrm{10 \ mHz}$,  $\mathrm{100 \ mHz}$, $\mathrm{200 \ mHz}$, $\mathrm{1 \ Hz}$) and amplitudes ($\mathrm{20\ mV_{pp}}$, $\mathrm{40\ mV_{pp}}$). The results indicate that the coupling is independent of the amplitude or frequency of the modulation. Table \ref{AS_to_IFO} provides an example of the measured coupling coefficients for every laser in the FSFBL-DFBL combination.

\begin{table}[htbp]
\centering
\begin{tabular}{cccc}
\toprule
     & \multicolumn{3}{c}{Coupling Factor (mrad/1)} \\
\cmidrule(lr){2-4}
Laser & DFBL-FBBL & FSFBL-DFBL & FSFBL-FBBL \\
\midrule
L1 & 2.40 & 0.65 & 2.87 \\
L2 & 1.95 & 2.50 & 3.70 \\
L3 & - & 0.24 & 0.10 \\
L4 & - & 0.14 & 0.09 \\
\bottomrule
\end{tabular}
\caption{Power-to-phase coupling coefficients for the individual lasers for each Backlink combinations.}\label{AS_to_IFO}
\end{table}

The coupling coefficients are dependent on the laser. The cause of this deviation is still unclear, but suggests that additional noise sources (e.g., polarization noise) may be involved. Now, we can estimate the actual power-to-phase coupling in our measurements. To illustrate this, figure \ref{fig:w_wo_AS} shows an example of a measurement between FSFBL and DFBL. Without active power stabilization (w/o PS), the projected total power noise (dashed blue) limits the measured differential non-reciprocity (solid green). The resulting noise exceeds the requirement, indicating that it is necessary to reduce these power fluctuations. For this purpose, we implemented a feedback control loop that adjusts and maintains the power through the respective VOAs. For each laser, a dedicated in-loop photodiode (PS1-PS4 in figure \ref{fig:3BLoptocad}) is installed on the optical bench to serve as a pick-off point to provide feedback signals. Now, with active power stabilization (w/ PS), the residual contribution of the projected total power noise (solid blue) is below the requirement across the entire measurement band. 

\subsubsection{Frequency Noise}
\label{sec:LFN_ifo}
Laser frequency noise (LFN) couples into the phase measurement when there is a mismatch in the optical pathlengths between the two interfering beams. The phase noise induced by LFN is \cite{Heinzel_2004}:
\begin{equation}
\begin{aligned} 
{\delta\varphi_{{\text{LFN}}}} &=   {\frac{2\pi\Delta L}{c}  {\delta f}},
\end{aligned}
\label{eq:LFN1}
\end{equation} 
where ${\Delta L}$ is the pathlength difference, $c$ is the speed of light, and $\delta f$ is the frequency noise, which is determined, in our case, by the reference laser.

In the 3BL, assuming an ideal case of mirror-symmetric optical benches, this contribution can be measured, together with the phase shift induced by the optical path of the Backlink, $\delta \varphi_\mathrm{L}$, when the Backlink interferometers in both benches are subtracted \cite{IsleifPhD}: 
\begin{align} 
\label{eq:sigma}
 \sigma_{\text{BL}}:= {\varphi^{\text{R}}_{{\text{BL}}}-\varphi^{\text{L}}_{{\text{BL}}}} &=  2\cdot[\delta \varphi_\text{LFN}+\delta \varphi_\text{L}] . 
\end{align}
The factor 2 accounts for the fact that, although these values are the same in both directions, they couple to the interferometer phase error with opposite signs on the two benches, therefore summing up in this measurement. In accordance, these contributions cancel when ${\varphi^{\text{L}}_{{\text{BL}}} \text{ and }\varphi^{\text{R}}_{{\text{BL}}}}$ are added, which corresponds to the non-reciprocity calculation shown in equation \ref{eq:singleBL}. However, under realistic conditions, a finite pathlength mismatch on the optical benches is unavoidable. For this reason, the 3BL was purposely set up with a dedicated frequency-stabilization system, which locks the reference laser to a hyperfine component of a molecular iodine transition, reducing $\delta \mathit {f}$ and, accordingly, $\mathrm{\delta\varphi_{{LFN}}}$ \cite{IsleifPhD}. The resulting frequency noise with active frequency stabilization (w/~FS), $\delta f^{\mathrm{w/}}_{\mathrm{FS}}$, is in the order of $\mathrm{100 \ \mathrm{Hz} /\sqrt{\text{Hz}}}$ at 0.1 Hz, increasing towards lower frequencies yet remaining under the requirement level across the spectrum \cite{MaikePhD}. 

\begin{figure}[h] 
\centering
\begin{subfigure}[t]{0.48\linewidth } \includegraphics[width=1\linewidth, trim=20 20 60 40, clip]{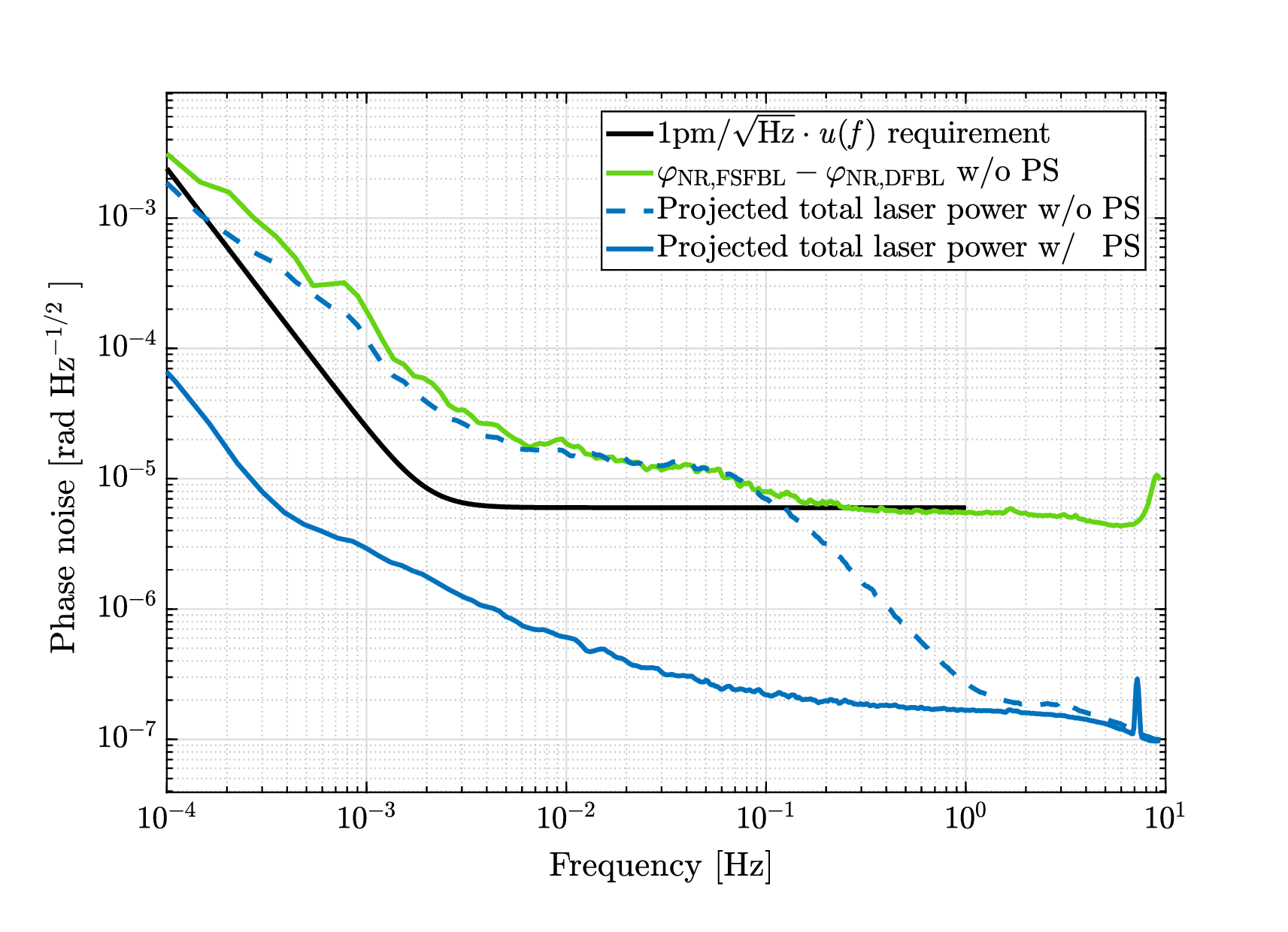}
\caption{Measured contribution of laser power noise into the differential non-reciprocity between DFBL and FSFBL. Without power stabilization (w/o PS), the projected total power noise (dashed blue) limits the measured differential non-reciprocity (solid green). With active power stabilization (w/ PS), the residual contribution of the projected total power noise (solid blue) is below the requirement across the entire measurement band.}
\label{fig:w_wo_AS} 
\end{subfigure}
\hfill 
\begin{subfigure}[t]{0.48\linewidth} \includegraphics[width=\linewidth, trim= 20 20 60 40, clip]{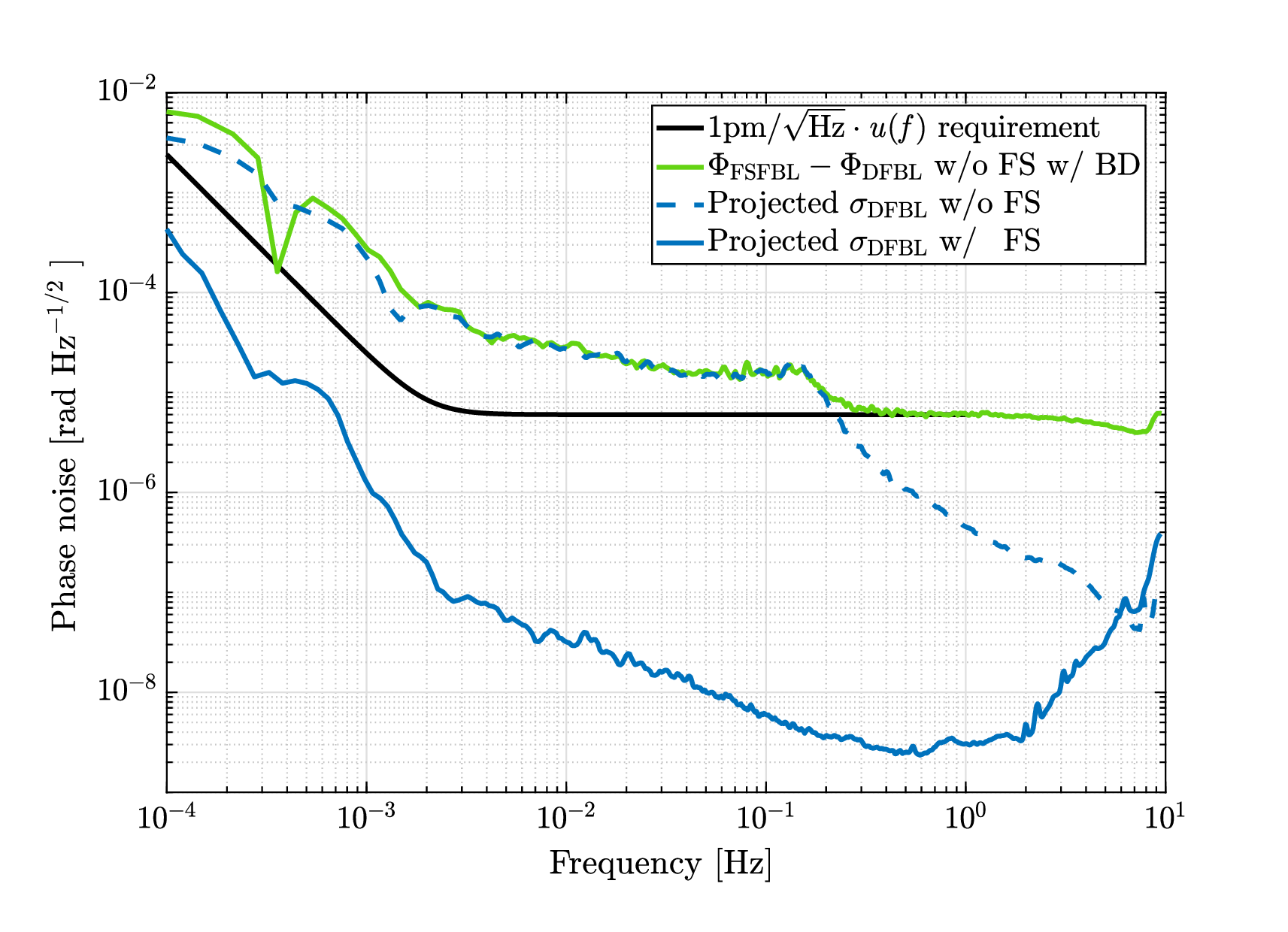}
\caption[LFN projection to NR]{Measured contributions of laser frequency noise into the differential non-reciprocity between DFBL and FSFBL with balanced detection (w/ BD). Without frequency stabilization (w/o FS), the $\sigma_{\mathrm{DFBL}}$ measurement (dashed blue) shows that LFN limits the differential non-reciprocity (solid green). With active frequency stabilization (w/ FS), the projected $\sigma_{\mathrm{DFBL}}$ (solid blue) is below the requirement across the entire measurement band.}    
\label{fig:w_wo_FS} 
\end{subfigure} 
\caption{Influence of laser power noise (a) and laser frequency noise (b) in the Three-Backlink Experiment.} 
\label{fig:laserNoise}
\end{figure}

Initial investigations of the coupling of LFN due to pathlength mismatches in the 3BL are reported in \cite{paperLea25,LeaBischofPhD}. However, due to a technical malfunction of the frequency stabilization system, the study could not be completed. Here, we present the follow-up investigation after the system was repaired. As explained in the initial study, when the additional pathlength mismatches in each Backlink implementation, $\mathrm{\Delta s}$, are considered in the differential non-reciprocity calculation, then equation (\ref{eq:nonRec}) becomes:  
\begin{equation}
{\Phi_{\text{BL1} }  - \Phi_{\text{BL2}}} = {\varphi_{\text{NR,BL1}} - {\varphi_{\text{NR,BL2}}}} +
 {\frac{2\pi \delta f} {c}} (\Delta s_1 - \Delta s_2 ).
\label{eq:NR_dS}
\end{equation}
Figure \ref{fig:w_wo_FS} illustrates an example to visualize this additional contribution. The differential non-reciprocity between FSFBL and DFBL (solid green) is obtained after applying balanced detection, as this is required to exclude coupling through backscatter, discussed in section \ref{sec:LFNcoupling}. It shows a shoulder-shape noise at $10^{-1}$~Hz, potentially originated from the LFN contribution due to asymmetries, bringing the noise above the requirement. This asymmetry is originated by the DFBL, as our observations indicate that it dominates over the other Backlink implementations. To estimate the magnitude of this asymmetry, we project the measurement obtained with equation \ref{eq:sigma} without active frequency stabilization (w/o FS) (dashed blue) into the differential non-reciprocity. The contribution of $\delta \varphi_\text{LFN}$ dominates for frequencies above $1.5~\text{mHz}$, and  $\delta \varphi_\text{L}$ for frequencies below. 

The frequency noise of the free-running laser was calculated, resulting in a worst-case value of $f^{\text{w/o}}_{\text{FS}}=17\ \text{kHz}/\sqrt{\text{Hz}} \cdot[1 \mathrm{Hz}/f]$. This is slightly above the typical value of $\delta f{_{\mathrm{NPRO}}} = 10 \ \mathrm{kHz}/ \sqrt{\text{Hz}} \cdot[1 \mathrm{Hz}/f]$, but the discrepancy matches with the literature \cite{Kwee_2012}. Now, we can determine the ratio between the projected $\sigma_{\mathrm{DFBL}}$ w/o FS and $\delta f^{\text{w/o}}_{\text{FS}}$, which results in $\Delta s_{\text{DFBL}} = 1.3$~mm following equation \ref{eq:LFN1}. Considering this worst-case mismatch, a frequency stability of at least  $\mathrm{221 \ \mathrm{kHz} /\sqrt{\text{Hz}}}\cdot u_\text{}(f)$ is necessary to fulfill the 1~\text{pm}/$\sqrt{\text{Hz}} \cdot u_\text{}(f)$ requirement. As previously mentioned, the current implementation reaches stability values of at least $\delta f^{\mathrm{w/}}_{\mathrm{FS}}=\mathrm{100 \ \mathrm{Hz} /\sqrt{\text{Hz}}}$, more than 3 orders of magnitude better than required. The resulting residual contribution of $\sigma_{\mathrm{DFBL}}$ w/~FS (solid blue) validates that the current frequency stabilization is sufficient, yielding the performance below the requirement.

\section{Backlink Noise}
\label{sec:backlinkNoise}
\smallskip

The noise inherent in the Backlink is determined by its design and configuration. Here, we examine the noises that have been identified so far, associated to the Backlink schemes that are integrated in the Three-Backlink Experiment: the fiber-based and the free-beam implementations. 

\subsection{Fiber-based Backlink implementation}
\label{sec:fiberbasednoise} 
To ensure that the implementation of a fiber-based Backlink is capable of reaching the requirement, it is essential to analyze how the non-reciprocal effects introduced by the fiber itself couple into the measurements. In this study, we focus only on the backscatter effect, which is a well-known limitation in the performance of a straightforward connection. In the 3BL, this concerns the DFBL implementation, as the FSFBL is designed to be insensitive to the effects of the fiber backscatter to the first order. Hence, in what follows, we will refer to the fiber-based Backlink implementation as simply DFBL.

\subsubsection{Backlink backscatter coupling} 

The dominant noise contribution in the DFBL implementation is expected to be the Rayleigh backscatter generated inside the Backlink fiber. The backscattered light will travel along with the counter-propagating Backlink nominal beam, entering the DFBL interferometer from the same port. This effect will introduce a phase error with opposite signs in the two interferometer outputs, thus enabling the possibility of suppressing this noise via balanced detection. Accordingly, in the contrary case where a differential phase measurement is performed between the two output ports, the phase noise contributions of the nominal beams will cancel, while the phase noise induced by backscatter will add constructively. We define this measurement as $\pi$-measurement, which proved to be a useful post-processing tool to investigate backscatter contributions in the individual interferometers.

The phase dynamics of the backscattered beam originated in the Backlink fiber are driven by the same factors as for the transmitted nominal beams. Hence, as described in section \ref{sec:LFN_ifo}, the main contributions are considered to be LFN coupling through the additional pathlength and temperature-induced changes in the Backlink fiber length.

In the following sections, we visualize this effect by considering the backscatter observed in the  $\pi$-measurement of the left DFBL interferometer: 

\begin{equation}
\label{eq:piMeasurement}
\delta\varphi^{\text{L}}_{\pi- \text{DFBL}} =\varphi^{\text{L\textsubscript{PD1}}}_{\text{DFBL}}-\varphi^{\text{L\textsubscript{PD2}}}_{\text{DFBL}}
\end{equation}
where $\varphi^{\text{L\textsubscript{PD1}}}_{\text{DFBL}}$ and $\varphi^{\text{L\textsubscript{PD2}}}_{\text{DFBL}}$ refer to the phase measured by the photodetectors of both output ports of the left DFBL interferometer. In addition, we consider two study cases, without and with frequency stabilization of the iodine reference laser. 

\begin{enumerate}[label=\roman*),ref=\roman*),ref=\thesubsubsection.\roman*),font=\itshape, leftmargin=0pt,itemindent=!,labelsep=.5em,widest=iv, listparindent=\parindent, parsep=0pt]
\item\label{sec:LFNcoupling} \textit{LFN coupling via Backlink backscatter}

For an unstabilized iodine reference laser, the Backlink-induced phase dynamics are dominated by LFN coupling in large parts of the measurement band. This effect was described in section \ref{sec:LFN_ifo}, where the noise coupling occurred via nominal asymmetries in the optical pathlengths. Accordingly, the results in figure \ref{fig:w_wo_FS} were obtained after applying balanced detection to suppress the coupling via backscatter, which is, in contrast, the object of study in the present section. The contribution of this effect to the non-reciprocity was already observed in \cite{LeaBischofPhD}. Here, we derive a linear coupling coefficient between the differential interferometer measurement, $\sigma_{\text{DFBL}}$ (equation \ref{eq:sigma}), and the left DFBL $\pi$-measurement, $\delta\varphi^{\text{L}}_{\pi- \text{DFBL}}$ (equation \ref{eq:piMeasurement}), by comparing the height of the shoulder-shaped LFN coupling contribution in both spectra. To enable a direct comparison, $\delta\varphi^{\text{L}}_{\pi- \text{DFBL}}$ is multiplied by a factor of $\tfrac{1}{2}$ to account for the fact that this observes twice the scatter in comparison to the nominal measurement of one output. The resulting coefficient will depend on the backscattered power and on the relative phase shift between the backscattered and nominal beams. To ensure averaging over all relative phase orientations, we considered a measurement segment with an additional, almost constant, temperature-induced phase drift. Figure \ref{fig:LFNtoPiCouplingASD} shows the ASD measured of half the $\pi$-measurement without frequency stabilization, $\frac{1}{2}\cdot \delta\varphi^{\text{L}}_{\pi- \text{DFBL}}$ w/o FS (solid green), corresponding to half the combined backscatter contribution of the two interferometer outputs. We observe that the projected differential interferometer noise without frequency stabilization, $\sigma_{\text{DFBL}}$ w/o FS (solid blue), reproduces very well the center-frequency shoulder, indicating that it is indeed dominated by LFN coupling. Now, if we use the same scaling factor to project the differential interferometer measurement with frequency stabilization, $\sigma_{\text{DFBL}}$ w/ FS (dash-dotted blue), we see that the significant improvement in frequency stability is expected to reduce the impact of LFN coupling through backscatter to below the required level. However, the $\pi$-measurement with frequency stabilization, $\frac{1}{2}\cdot \delta\varphi^{\text{L}}_{\pi- \text{DFBL}}$ w/ FS (dash-dotted green), does not show the same level of improvement. In our current understanding, the remaining shoulder between 1\,mHz and 50\,mHz is not related to Backlink backscatter, as discussed in section \ref{sec:results}. In addition, the deviation from the $\pi$-measurement in the frequency range below approximately 1\,mHz results from a non-linear coupling of temperature-induced length changes of the Backlink fiber. This effect is described in the following section \ref{sec:T2Pscatter}.
\begin{figure}[ht] 
\centering
\begin{subfigure}[t]{0.49\linewidth } \includegraphics[width=1\linewidth,
trim= 20 25 20 30, clip]
{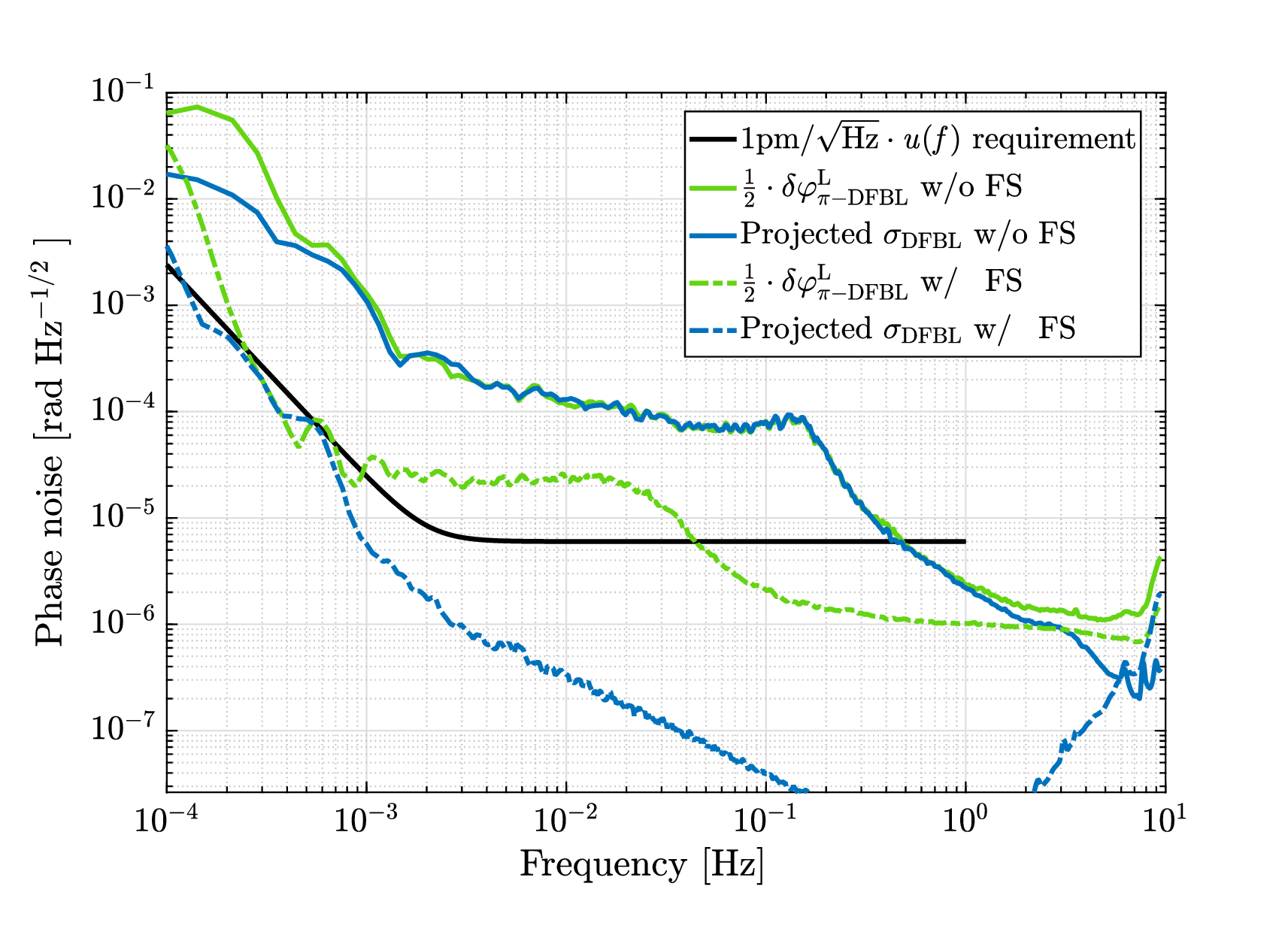}
\caption{Laser frequency noise coupling via Backlink backscatter.}
\label{fig:LFNtoPiCouplingASD} 
\end{subfigure}
\begin{subfigure}[t]{0.5\linewidth} \includegraphics[width=\linewidth, trim= 30 35 10 30, clip]{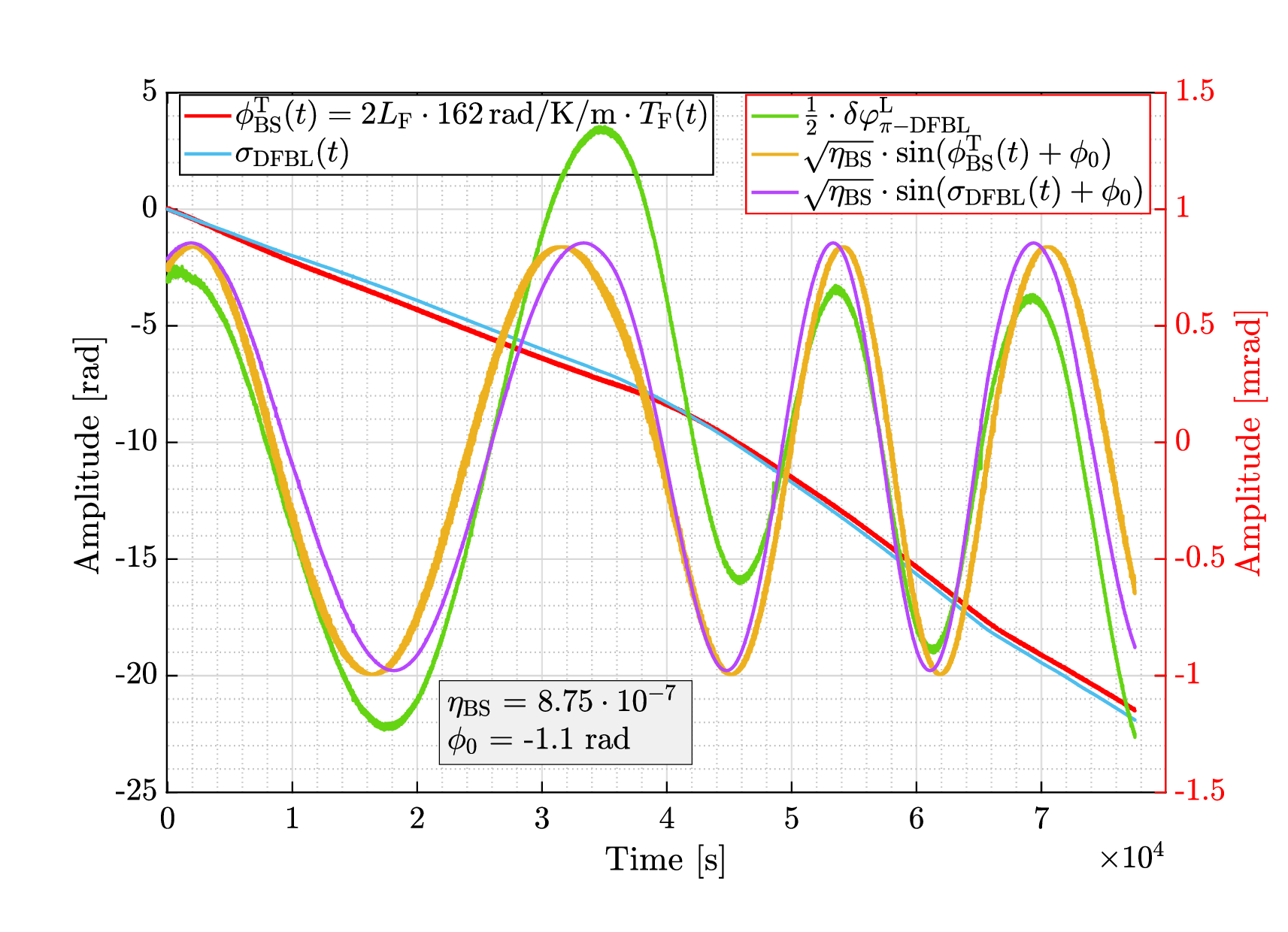}\caption{Non-linear backscatter coupling from temperature drifts.}\label{fig:DriftCoupling}\end{subfigure} 
\caption{Measurement of the coupling between backscatter and phase noise, originated from laser frequency noise (a) and temperature variations (b). The values of $\delta\varphi^{\text{L}}_{\pi- \text{DFBL}}$ (solid, dashed green),  $\sigma_{\text{DFBL}}$ (solid, dashed blue), and $T_{\mathrm{F}}(t)$ are measured experimentally, which are included in $\phi_\text{BS}^\text{T}(t)$ (solid red) and the two fits of $\delta\varphi^{\text{DFBL}}_\text{BS}$ (solid orange and purple), which follow equation \ref{eq:BL_BS_Atan}.}
\label{fig:BackscatterCoupling}
\end{figure}

\item\label{sec:T2Pscatter} 
\textit{Temperature-to-phase coupling via Backlink backscatter}

Temperature fluctuations induce variations in the length of the Backlink fiber, leading to phase shifts of the backscattered light. Measurements of the temperature-to-phase coupling, ${\tfrac{\delta\upphi}{\delta T}}$, of the Nufern PM1060L fiber, which is comparable to the Nufern PM980-XP fiber of the 3BL, show values ranging from 35--50\,rad/K/m \cite{RohrPhD} to 77\,rad/K/m  \cite{Campbell} per unit of fiber length. These results were obtained from fibers without a jacket and are potentially higher for the 3BL fibers, as they feature an additional wrapping in FEP (fluorinated ethylene propylene). Our current measurements of the 3BL indicate a temperature-to-phase coupling in the order of $[170\pm10]$\,rad/K/m. However, unlike the cited references, the coupling was not obtained from a dedicated temperature injection on the Backlink fiber. Instead, we deduced it by scaling the temperature data measured close to the Backlink fiber (figure \ref{fig:DriftCoupling}, red, left y-axis) to the phase shift induced by the backlink path (figure \ref{fig:DriftCoupling}, blue, left y-axis), measured in the differential interferometer combination, $\sigma_{\text{DFBL}}$.

A linearized model for the backscatter phase error induced by optical length changes of the Backlink is given in \cite{LISA-LCST-INST-TN-003}. This linearized description only captures effects of in-band temperature fluctuations which are considered small. However, in the 3BL, we observe a non-linear coupling of lower frequency temperature drifts, creating the need of a more generalized approach.  
The impact of Backlink backscatter on the interferometer phase error, defined as $\delta\varphi^{\text{BL}}_\text{BS}$, will depend on both the power ratio, $\eta$, and the relative phase shift of the backscattered beam, $\phi_\text{BS}(t)$, with respect to the local reference beam that enters the interferometer from the other port. Assuming perfect polarization and mode overlap, a straightforward combination of the complex field vectors between the backscatter and nominal interferometric signal results in the following phase error: 
\begin{equation}
\label{eq:BL_BS_Atan}
\delta\varphi^{\text{BL}}_\text{BS} = \arctan\left( \frac{\sin(\phi_\text{BS}(t))}{\tfrac{1}{\sqrt{\eta}}+\cos(\phi_\text{BS}(t))}\right) \stackrel{\tfrac{1}{\sqrt{\eta}}\gg1}{\approx} \sqrt{\eta}\cdot\sin(\phi_\text{BS}(t))
\end{equation}
 The approximation assumes that the amplitude of the backscatter is small compared to the amplitude of the nominal beam. In agreement with the model in \cite{LISA-LCST-INST-TN-003}, the temperature-induced phase shift for backscatter that experiences twice the Backlink length $L_\text{F}$ (worst case) in a temperature environment $\delta T(t)$ can be described by $\phi_\text{BS}^\text{T}(t)=\tfrac{\delta\phi}{\delta T} \cdot 2L_\text{F}  \cdot \delta T(t)$. A realistic example for this scenario would be e.g. backscatter from the Backlink outcoupling FIOS propagating back through the complete fiber length. 

According to equation~\ref{eq:BL_BS_Atan}, if a temperature drift shifts the backscatter phase over multiple radians, this will be converted into an oscillation of the interferometer phase error, as illustrated with the $\pi$-measurement (green, right y-axis) in figure \ref{fig:DriftCoupling}. For comparison, the model of equation \ref{eq:BL_BS_Atan} is shown as well. The backscatter phase, $\phi_\text{BS}(t)$, uses the scaled temperature data,  $\phi_\text{BS}^\text{T}(t)$ (orange, left y-axis), in one case, and the differential interferometer data, $\sigma_{\text{DFBL}}$ (purple, right y-axis), in the other, with an additional offset $\phi_0$. We obtained a power ratio of { $\eta=8.75\cdot10^{-7}$} and an offset phase shift of $\phi_0=-1.1$ rad by applying a least squares fit for the model using differential interferometer data, $\sigma_{\text{DFBL}}$. Though the model does not fully describe the measured backscatter in the $\pi$-measurement, it qualitatively reproduces the impact of the temperature drift. The frequency of the induced oscillation depends linearly on the temperature slope. This way, out-of-band drifts can be converted into in-band phase noise. In section~\ref{sec:results}, the ASD of the model is shown, which will explain the low-frequency noise in the measurements without balanced detection.

\end{enumerate}

\subsection{Free-beam Backlink implementation }

In a Backlink transmitted in free-space, the angular jitter of the beam is a critical parameter that needs to be minimized to guarantee that the performance meets the requirement. In the 3BL, this concerns the FBBL, which uses an active control loop to correct the beam tilt while maintaining the link when the benches are rotated. Here, we describe the integration and performance of this implementation. 

%
\subsubsection{DWS control loop}\label{sec:DWS} 
The FBBL uses two active steering mirrors to transmit the beams between the benches. To maintain the interferometric contrast, it uses a control loop based on Differential Wavefront Sensing (DWS) \cite{paperGerhard2020}, which senses four angles on two quadrant photodetectors (QPDs). The signals of each segment of the QPD are sent to the readout Phasemeter, where a dedicated control loop is implemented to keep the alignment by sending a high-voltage electric signal to the piezo actuators of the two steering mirrors \cite{JestrabekMaster, BischofMaster2018}. Due to degradation over time, the piezo actuators have been replaced with respect to these references. The E-616 by Physik Instrumente have been switched to the PSH 25 by Piezosystem Jena. This improved the performance of the control system as a result of the finer angular resolution and higher bandwidth of the latter. Additionally, precise optimization of the feedback control loop parameters is essential to achieve rapid and effective correction of the dynamic fluctuations of the signals. To ensure that the beam tilt does not exceed the 1\text{\ pm}/$\sqrt{\text{Hz}} \cdot u_\text{}(f)$ pathlength noise requirement, an additional requirement for the DWS was derived in \cite{IsleifPhD}. Additionally, \cite{LeaBischofPhD} provides the exact value of the tilt-to-length (TTL) misalignment in the FBBL after construction. Considering a TTL coupling coefficient of 139\mbox{\ $\mathrm{pm}/\mathrm{\upmu rad}$} and a DWS coupling coefficient (CC) of 3024 $\mathrm{rad}/\mathrm{rad}$, obtained after the control loop implementation in the system with the new piezo actuators, the DWS requirement becomes:
\begin{equation}
\mathrm{DWS}_{\mathrm{req}} = 1~\mathrm{pm} / \sqrt{\mathrm{Hz}} \cdot u_\text{}(f)
 \cdot \frac{\mathrm{CC}}{\mathrm{TTL}}   \simeq 2.2\cdot10^{-5}\frac{\mathrm{rad}}{\sqrt{\mathrm{Hz}}} \cdot u_\text{}(f).
\label{eq:DWSreq}
\end{equation}
Consequently, this value determines a coupling factor of the DWS noise to the non-reciprocity measurements, as shown in figures \ref{fig:nonrec_DFBL_FBBL} and \ref{fig:nonrec_FBBL_FSFBL}.

\section{Results}
\label{sec:results}
\smallskip

This section presents the current performance of the Three-Backlink Experiment. The differential non-reciprocal phase noise of each Backlink combination is illustrated separately in figures \ref{fig:nonrec_DFBL_FBBL}-\ref{fig:nonrec_FBBL_FSFBL}. Each plot includes the 1~\text{pm}/$\sqrt{\text{Hz}} \cdot u_\text{}(f)$ requirement (solid black), a 3~\text{pm}/$\sqrt{\text{Hz}} \cdot u_\text{}(f)$ relaxed requirement for reference (dashed-dotted gray), and the differential non-reciprocity measurements with (solid orange) and without balanced detection (dash-dotted orange), both with active laser power and frequency stabilization. The noise sources described in sections \ref{sec:commissioning} and \ref{sec:backlinkNoise} are quantified for each case, and the resulting total noise budget (yellow shaded area), which assumes all noises to be uncorrelated and balanced detection, is limited by the best- and worst-case noise predictions. The upper (solid light blue) and lower (solid dark blue) limits of the readout noise, which includes the Phasemeter and TIA noises, are derived from repeated dark-noise measurements of the electronics at the operating beatnote frequencies for each combination, and subsequently summed according to the number of times each frequency appears in the measurement. Additionally, the detector noise (solid light green), which includes RIN, SN and EN, is depicted as a summed contribution for simplicity; the relative weights are different for every case and the limiting one will be specified in the corresponding section. For the laser noise, the power noise (solid pink) includes the total contribution of the lasers, and the frequency noise (solid red) represents the coupling via asymmetry from the data of a stabilized iodine reference, provided by \cite{VictorPhD}. Regarding the noise inherent to the Backlink designs, we include the DWS (solid dark green) for the free-space implementation, and the backscatter coupling for the fiber-based DFBL implementation, both with (solid purple) and without (dash-dotted purple) balanced detection. Each plot includes an inset that represents the environmental temperature noise in the laboratory, which has been measured simultaneously with the differential non-reciprocity. The inset shares the x-axis with the main plot and uses different y-axis units. The aspect ratio between the plots is matched to enable a direct visual comparison of the curves. The lower and upper limits of the total noise budget at low frequencies were derived from the temperature fluctuations with worst or better temperature stability in the laboratory. Finally, for clarity, we divide the measurement frequency band into three ranges (vertical dashed gray lines), corresponding to regimes in which different noise sources dominate. These are the high-frequency
range for $8\cdot 10^{-2} \ \text{Hz} < f $, the center-frequency range for $1.5\cdot 10^{-3} \ \text{Hz} < f < 8\cdot 10^{-2} \ \text{Hz} $, and the
low-frequency range for $f< 1.5\cdot 10^{-3} \ \text{Hz}$.

\subsection{Direct Fiber Backlink vs Free-Beam Backlink}
\label{sec:DFBL_FBBL}

\begin{figure}[ht]
    \centering
     \includegraphics[width=1.0\linewidth, trim=40 20 60 40, clip]{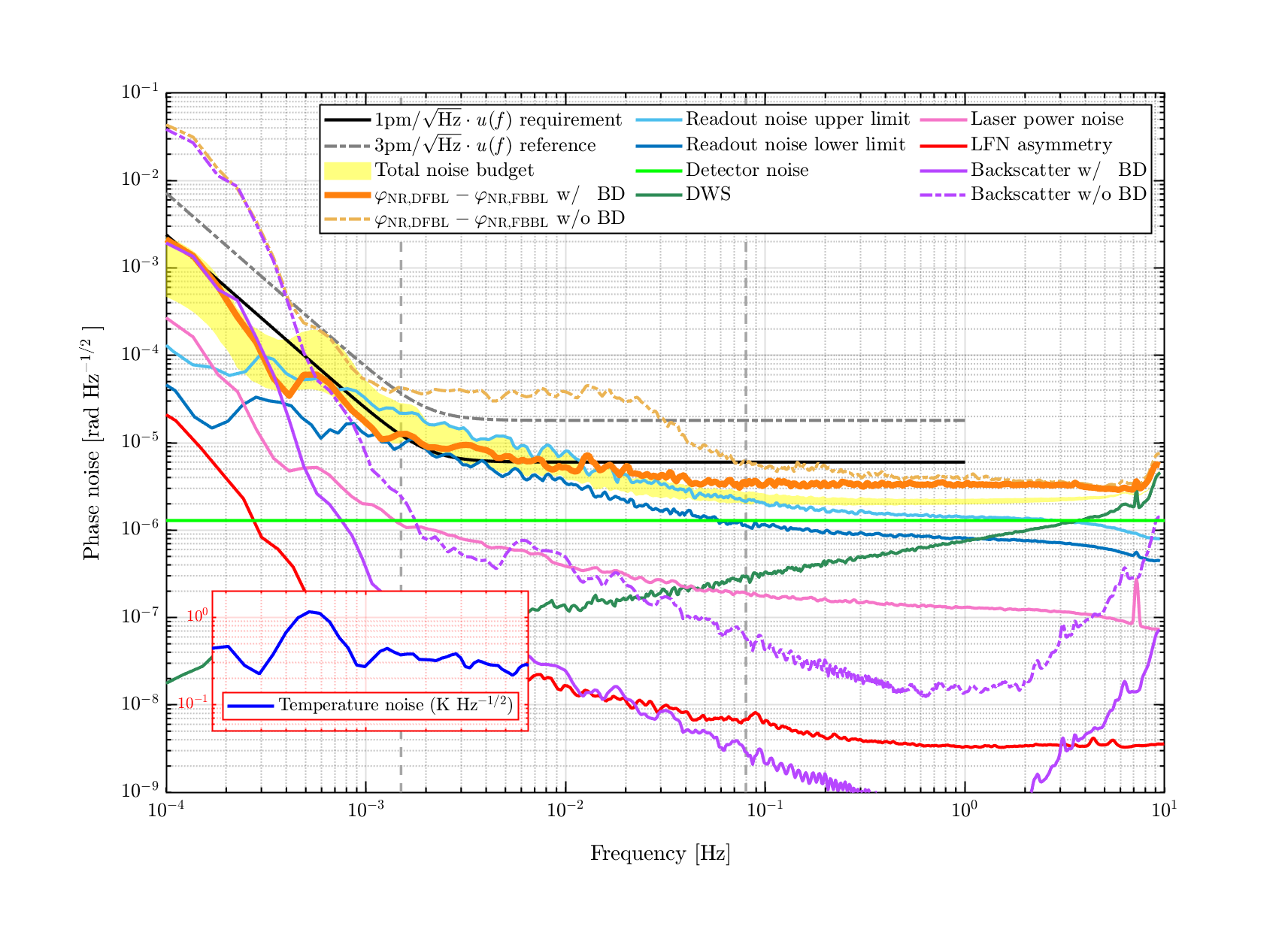 }
    \caption[Differential non-reciprocity between DFBL-FBBL]{Phase noise performance between DFBL and FBBL. The frequency span is divided into \mbox{low-,} \mbox{center-,} and high-frequency ranges for visual clarity, indicated by vertical dashed-gray lines. The individual noise contributions are assumed to be uncorrelated and summed in quadrature, resulting in the predicted total noise budget (yellow shaded area). The figure shows the differential non-reciprocities with (solid orange) and without (dash-dotted orange) balanced detection (BD). The differential non-reciprocity with BD meets the 1~\text{pm}/$\sqrt{\text{Hz}} \cdot u_\text{}(f)$ requirement (solid black) across the entire measurement band, with a marginal exceedance between 2 mHz and 4 mHz due to the testbed noise limitations, and remains below 3~\text{pm}/$\sqrt{\text{Hz}} \cdot u_\text{}(f)$ (dash-dotted gray) over the entire span.} 
    \label{fig:nonrec_DFBL_FBBL}    
\end{figure}

A key advantage of measuring the differential non-reciprocity between DFBL and FBBL, \linebreak $\mathrm{\Phi_{DFBL}} $ $ \mathrm{- \ \Phi_{FBBL}}$, is that both designs require only two lasers, which simplifies the setup and reduces potential points of failure. As it operates with only one beatnote, it is possible to effectively reduce the noise originated by the readout electronics by choosing a signal at a lower frequency (see section \ref{sec:detectionNoise}). The current performance of this combination is shown in figure \ref{fig:nonrec_DFBL_FBBL}, and the following analysis evaluates the non-reciprocal phase noise obtained with and without balanced detection.

\subsubsection{With balanced detection} 
The measured differential non-reciprocity with balanced detection, $\mathrm{\varphi_{NR,DFBL} - {\varphi_{NR,FBBL}}}$ w/ BD, meets the 1~\text{pm}/$\sqrt{\text{Hz}} \cdot u_\text{}(f)$ requirement across the entire measurement band, with a marginal exceedance between 2 mHz and 4 mHz due to the testbed noise limitations, thus not related to the Backlink designs. Nevertheless, it remains below  3~\text{pm}/$\sqrt{\text{Hz}} \cdot u_\text{}(f)$ over the entire span.

The high-frequency range is mainly limited by the detector noise, in this case particularly shot noise, and the readout noise. At frequencies beyond the measurement bandwidth, the contribution of the DWS control loop gains influence. The measured differential non-reciprocity exhibits a noise floor of 0.57 \text{pm}/$\sqrt{\text{Hz}}$ ($\mathrm{3.35 \  \mathrm{\upmu}rad /\sqrt{\text{Hz}}}$), which is slightly above the prediction but below the requirement by a factor of 1.75. 

The center-frequency range is also limited by the detector and readout noise contributions, with the latter being more relevant in this case. Its upper and lower limits define the performance boundaries for a beatnote frequency of 212.119 kHz. The differential non-reciprocity measurement slightly exceeds the requirement between 2 mHz and 4 mHz, reaching a maximum value of 1.48~\text{pm}/$\sqrt{\text{Hz}}\cdot u_\text{}(f)$ ($\mathrm{8.77 \ \mathrm{\upmu}rad /\sqrt{\text{Hz}}}\cdot u_\text{}(f)$). However, this lies within the predicted noise budget, as the best-case estimate is already on the 1 \text{pm}/$\sqrt{\text{Hz}}$ level.

The low-frequency range is limited by the temperature-to-phase coupling via backscatter in the Backlink, and potentially by temperature fluctuations in the laboratory environment. While the coupling mechanism of the environmental temperature fluctuations is still under investigation (see section \ref{sec:EnvTemp}), this assumption is corroborated by the fact that both the measured laboratory temperature noise and the non-reciprocity feature the same bump at 0.6 mHz, which corresponds to the air-conditioning cycle of the HVAC. The effect of the temperature-to-phase coupling via backscatter (see section \ref{sec:T2Pscatter} is visible in the lower noise shoulder, as confirmed by the close overlap between the backscatter model and the measurement. The coupling occurs presumably for two reasons. First, the temperature drift during the time of this measurement, resulting in the in-band bump and, second, the balanced detection efficiency was 97\%, suggesting the influence of the remaining contribution. 

\subsubsection{Without balanced detection}
The measured differential non-reciprocity without balanced detection, $\mathrm{\varphi_{NR,DFBL} - {\varphi_{NR,FBBL}}}$ w/o BD, meets the requirement only in the high-frequency range. In the center-frequency range, the shoulder-shaped noise suggests that it is limited by spurious straylight of the optical bench (see section \ref{sec:straylight}). The same noise appears in both the DFBL and FBBL, suggesting that it is unlikely that it originates from the DFBL fiber. In the low-frequency range, the increased noise is due to the temperature-to-phase coupling via backscatter (see section \ref{sec:T2Pscatter}). As explained in the previous paragraph, the temperature drift was higher than the threshold, resulting in the in-band coupling. The modeled backscatter noise w/o BD shows exactly this effect.

\subsection{Frequency-Separated Fiber Backlink vs Direct Fiber Backlink}
\label{sec:FSFBL-DFBL}
\begin{figure}[ht]
    \centering
    \includegraphics[width=1.0\linewidth, trim= 40 20 60 40, clip]{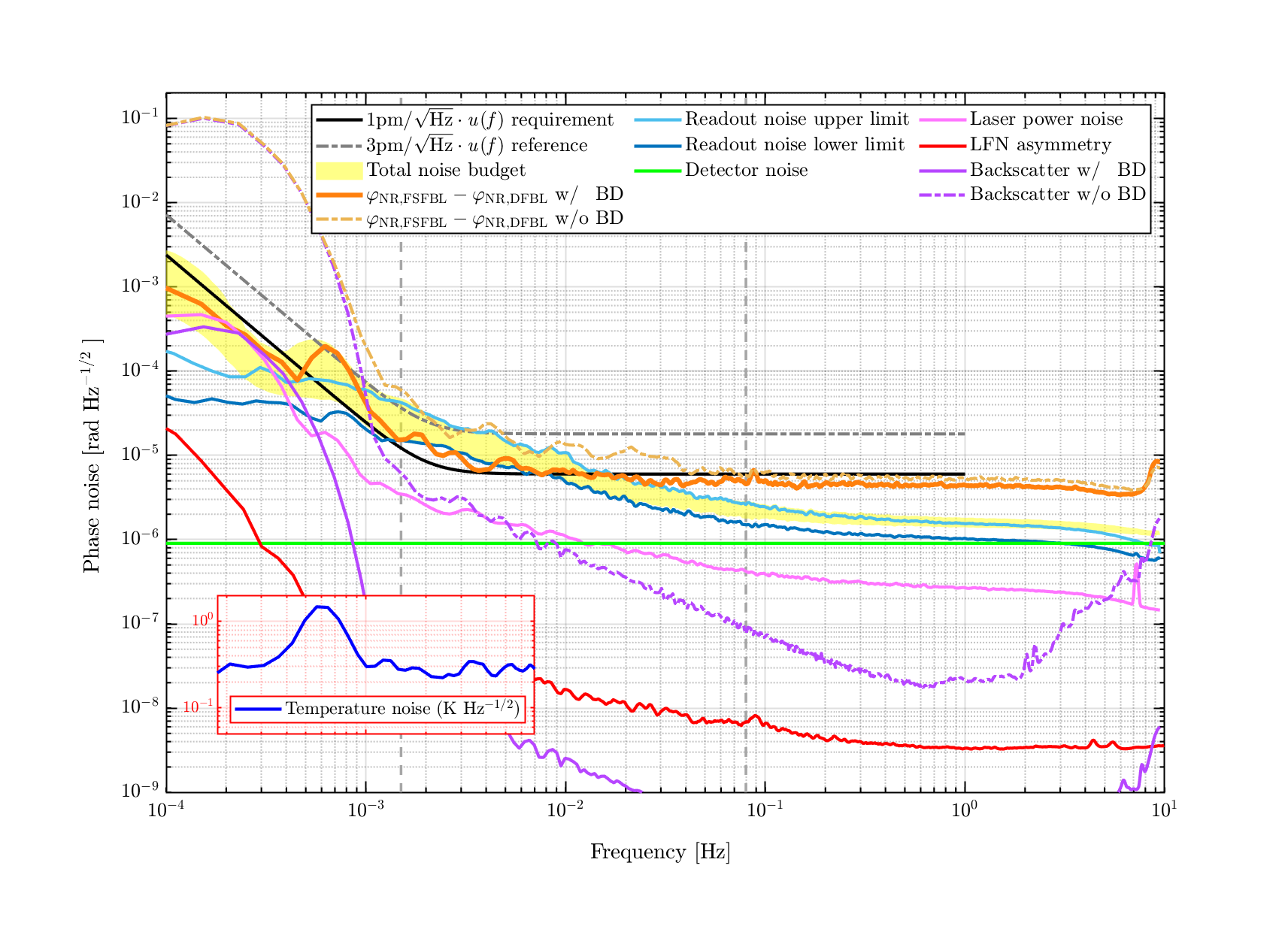 }
       
    \caption[Differential non-reciprocity between FSFBL-DFBL]{Phase noise performance between FSFBL and DFBL. The frequency span is divided into \mbox{low-,} \mbox{center-,} and high-frequency ranges for visual clarity, indicated by vertical dashed-gray lines. The individual noise contributions are assumed to be uncorrelated and summed in quadrature, resulting in the predicted total noise budget (yellow shaded area). The figure shows the differential non-reciprocities with (solid orange) and without (dash-dotted orange) balanced detection (BD). The differential non-reciprocity with BD meets the 1~\text{pm}/$\sqrt{\text{Hz}} \cdot u_\text{}(f)$ requirement (solid black) across most of the measurement band, except from $0.4 \text{ mHz} $ to $6 \text{ mHz}$ due to the testbed noise limitations, and remains below  3~\text{pm}/$\sqrt{\text{Hz}} \cdot u_\text{}(f)$ (dash-dotted gray) over the entire span. }

    \label{fig:nonrec_FSFBL_DFBL}
\end{figure}

The differential non-reciprocity measurement between FSFBL and DFBL, $\mathrm{\Phi_{FSFBL}  - \Phi_{DFBL}}$, allows the study of the phase noise originated from fiber-based Backlink designs, excluding any noise derived from a free-beam scheme. The current performance of this combination is shown in figure \ref{fig:nonrec_FSFBL_DFBL}, and the following analysis evaluates the non-reciprocal phase noise obtained with and without balanced detection.

\subsubsection{With balanced detection}
The measured differential non-reciprocity with balanced detection, $\mathrm{\varphi_{NR,FSFBL} - {\varphi_{NR,DFBL}}}$ w/ BD, meets the 1~\text{pm}/$\sqrt{\text{Hz}} \cdot u_\text{}(f)$ requirement across most of the measurement band, exceeding it between $0.4 \text{ mHz} $ to $6 \text{ mHz}$ due to the testbed noise limitations, thus not related to the Backlink designs. Nevertheless, it remains below  3~\text{pm}/$\sqrt{\text{Hz}} \cdot u_\text{}(f)$ over the entire span.

The high-frequency range is limited by the detector noise, in this case electronic noise, and readout noise. 
The measured differential non-reciprocity exhibits a noise floor of 0.74~\text{pm}/$\sqrt{\text{Hz}}$ \ ($\mathrm{4.38 \  \mathrm{\upmu}rad /\sqrt{\text{Hz}}}$), which is roughly a factor of 2 above the predicted noise and a factor of 1.35 below the requirement. 

The center-frequency range is mainly constrained by the readout noise. The upper and lower limit curves are obtained by combining the dark-noise measurements at the different frequencies present in this combination, [DFBL -- FSFBL -- REF], which are [212.119 -- 231.054 -- 248.345] kHz on the left bench and  [212.119 -- 460.464 -- 443.173] kHz on the right bench. As explained in section \ref{sec:readoutNoise_LF}, the readout electronics exhibits an upper boundary due to signal attenuation at frequencies near 500 kHz. Since this combination involves frequencies close to this limit, the overall level increases accordingly. Already the best-case estimation exceeds the requirement at frequencies between 1 mHz and 6 mHz, meaning that it is technically not possible to reach performance.  The highest measured phase noise is 1.84 \text{pm}/$\sqrt{\text{Hz}} \cdot u_\text{}(f)$ ($\mathrm{10.9\  \mathrm{\upmu}rad /\sqrt{\text{Hz}}} \cdot u_\text{}(f)$), which agrees with the estimated noise budget.

The low-frequency range is potentially limited by temperature fluctuations in the laboratory environment. As in section \ref{sec:DFBL_FBBL}, this measurement also features the noise bump at 0.6 mHz corresponding to the air-conditioning cycle of the HVAC. However, since more readout channels are used in this combination due to the additional REF interferometers, the noise is added accordingly, reaching the 3 \text{pm}/$\sqrt{\text{Hz}}\cdot u_\text{}(f)$ level. On the other hand, the lower bump originated from the temperature-to-phase coupling via backscatter in the DFBL fiber seems to not be the limiting noise in this case. The balanced detection efficiency for this measurements was higher, and the shown modeled backscatter curve represents an upper limit. 

\subsubsection{Without balanced detection}
The measured differential non-reciprocity without balanced detection, $\mathrm{\varphi_{NR,FSFBL} - {\varphi_{NR,DFBL}}}$ w/o BD, meets the requirement only in the high-frequency range. As in the previous section, the DFBL is limited by spurious straylight of the optical bench in the center-frequency range ---which features a lower shoulder as the contribution of FBBL is not included in this case--, and by the temperature-to-phase coupling via backscatter in the Backlink fiber in the low-frequency range, well explained with the modeled backscatter noise w/o BD. The FSFBL does not see any improvement with balanced detection, which is consistent with this implementation, as it is purposely designed to prevent these effects with the use of separate signal frequencies.

\subsection{Free-Beam Backlink vs Frequency-Separated Fiber Backlink}

\begin{figure}[ht]
    \centering
    \includegraphics[width=1.0\linewidth, trim= 40 20 60 40, clip]{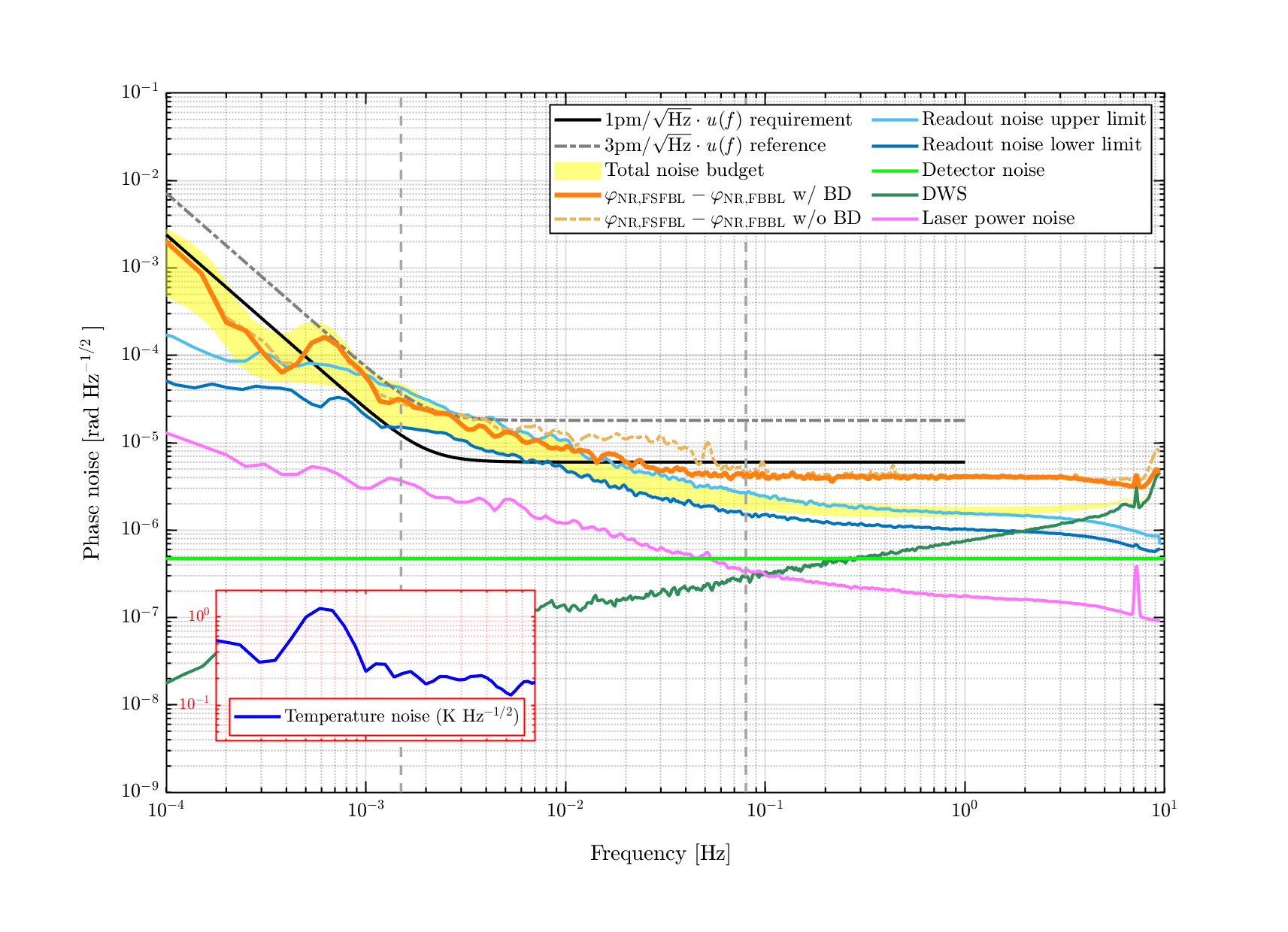 }
     \caption[Differential non-reciprocity between FSFBL-FBBL]{Phase noise performance between FSFBL and FBBL. The frequency span is divided into \mbox{low-,} \mbox{center-,} and high-frequency ranges for visual clarity, indicated by vertical dashed-gray lines. The individual noise contributions are assumed to be uncorrelated and summed in quadrature, resulting in the predicted total noise budget (yellow shaded area). The figure shows the differential non-reciprocities with (solid orange) and without (dash-dotted orange) balanced detection (BD). The differential non-reciprocity with BD meets the 1~\text{pm}/$\sqrt{\text{Hz}} \cdot u_\text{}(f)$ requirement (solid black), except from 0.5 mHz to 20 mHz due to the testbed noise limitations, and remains below 3~\text{pm}/$\sqrt{\text{Hz}} \cdot u_\text{}(f)$ (dash-dotted gray) over the entire span.}     
    \label{fig:nonrec_FBBL_FSFBL}
\end{figure}

The differential non-reciprocity combination between FBBL and FSFBL, $\mathrm{\Phi_{FBBL}  - \Phi_{FSFBL}}$, allows the possibility to study the performance of two Backlink designs that overcome the fiber backscatter problem, which is inherent in a straightforward Backlink scheme like the DFBL. The current performance of this combination is shown in figure \ref{fig:nonrec_FBBL_FSFBL}, and the following analysis evaluates the non-reciprocal phase noise obtained with and without balanced detection.
\subsubsection{With balanced detection}
The measured differential non-reciprocity with balanced detection, $\mathrm{\varphi_{NR,FBBL} - {\varphi_{NR,FSFBL}}}$ w/ BD, meets the 1~\text{pm}/$\sqrt{\text{Hz}} \cdot u_\text{}(f)$ requirement outside the $0.5\text{ mHz} -20\text{ mHz}$ range, being limited by testbed noise limitations, thus not related to the Backlink designs. Nevertheless, it remains below the  3~\text{pm}/$\sqrt{\text{Hz}} \cdot u_\text{}(f)$ reference within the entire span. The projection of LFN coupling via asymmetry is omitted because it could not be determined for the FSFBL and FFBL schemes, as any contribution was already limited by the noise floor. However, this likely implies that its contribution is negligible. 

The high-frequency range is mostly limited by the readout noise. The detector noise, which is the lowest in this combination, is dominated by electronic noise. The measured differential non-reciprocity is 0.68~\text{pm}/$\sqrt{\text{Hz}}$ ($\mathrm{4.0 \  \mathrm{\upmu}rad /\sqrt{\text{Hz}}}$), which is a factor of 2 above the predicted noise and a factor of 1.5 below the requirement. 

The center-frequency range is dominated by the readout noise due to the FSFBL design, as explained in section \ref{sec:FSFBL-DFBL}. However, in this case the performance shows a steeper slope towards lower frequencies, suggesting a higher coupling of the noise. This results in a worst performance, which exceeds the requirement at a slightly earlier frequency of about 20 mHz, while still lying within the predicted boundaries.

The low-frequency range is limited by temperature fluctuations in the laboratory environment. As this combination is analogous to the FSFBL-DFBL case, the noise explanation is omitted to avoid repetition. 

\hspace{1pt}

\subsubsection{Without balanced detection}
The measured differential non-reciprocity without balanced detection, $\mathrm{\varphi_{NR,FBBL} - {\varphi_{NR,FSFBL}}}$ w/o BD, overlaps with the measurement with balanced detection in the high- and low-frequency ranges. In the center-frequency range, the shoulder-shaped noise corresponds to the straylight coupled into the FBBL implementation, which is non-intrinsic to the Backlink design (see section \ref{sec:straylight}). Future work is planned to mitigate this effect, aiming to achieve the desired performance without the need for balanced detection. Overall, this measurement confirms that the FSFBL implementation, although being a fiber-based Backlink, is insensitive to the fiber backscatter by the design choice of using separated frequencies in the signals. This is particularly visible with the overlap of the two measurements --with and without balanced detection-- at low frequencies. 

\subsection{Key Findings and Implications}
\label{sec:key}

\begin{figure}[h]
    \centering
   \includegraphics[width=0.9\linewidth, trim= 20 20 60 40, clip]
    {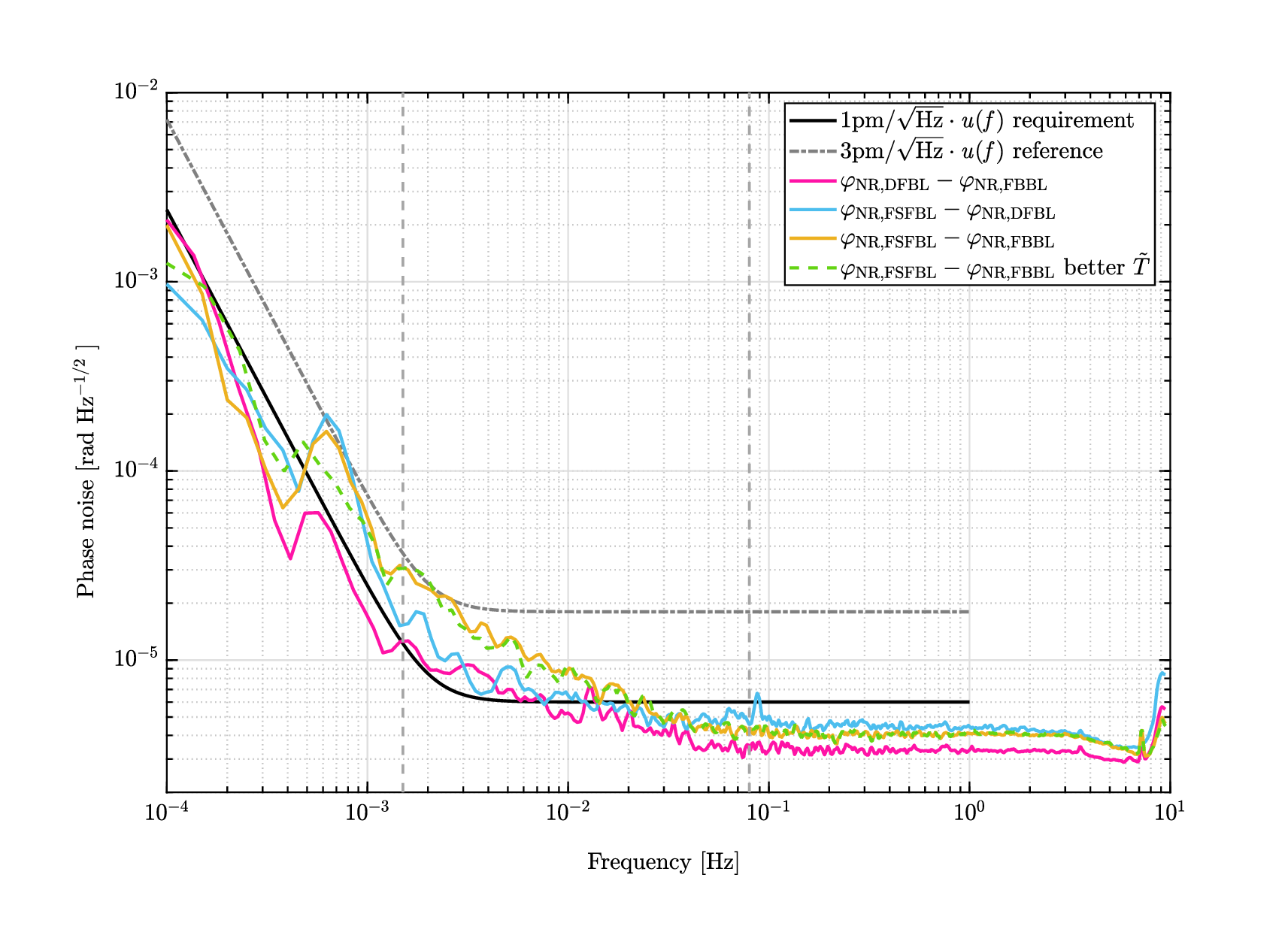}
    \caption[Current performance 3BL]{Current performance of the Three-Backlink Experiment. The frequency span is divided into \mbox{low-,} \mbox{center-,} and high-frequency ranges for visual clarity, indicated by vertical dashed-gray lines. The figure illustrates the performances with balanced detection of the three possible Backlink combinations: DFBL-FBBL (solid magenta), FSFBL-DFBL (solid blue), and FSFBL-FBBL (solid orange). For the FBBL-FSFBL combination, a segment with better temperature stability, $\tilde{T}$, is included (dashed green) to illustrate the reduction of the peak at 0.6 mHz. All differential non-reciprocities meet the 1~\text{pm}/$\sqrt{\text{Hz}} \cdot u_\text{}(f)$ requirement (solid black) across most of the frequency band, being limited by the respective testbed noise in each combination, and remain below 3~\text{pm}/$\sqrt{\text{Hz}} \cdot u_\text{}(f)$ (dash-dotted gray) over the entire span.}
    \label{fig:nonrec_performance}
\end{figure}

The current performance of the Three-Backlink Experiment is illustrated in figure \ref{fig:nonrec_performance}, which contains the three differential non-reciprocity combinations: $\mathrm{\Phi_{DFBL}}- \mathrm{\Phi_{FBBL}}$ (solid magenta), $\mathrm{\Phi_{FSFBL}}- \mathrm{\Phi_{DFBL}}$ (solid blue), and $\mathrm{\Phi_{FSFBL}}- \mathrm{\Phi_{FBBL}}$ (solid orange). Although these measurements are not obtained simultaneously --due to the limited number of channels in the Phasemeter--, they are presented together to facilitate visual comparison. The results show that all Backlink implementations perform below the $3 \ \text{pm}/\sqrt{\text{Hz}} \cdot ~u_\text{}(f)$ reference (dash-dotted gray) across the entire measurement band, and below the $1 \ \text{pm}/\sqrt{\text{Hz}} \cdot ~u_\text{}(f)$ requirement (solid black) across most of it. The sensitivity is exceeded between 0.5 mHz$-$20 mHz in the worst case, and between 2 mHz$-$4 mHz in the best case. The origin of this excess noise is extrinsic to the Backlink designs, and due to technical limitations in the readout electronics and the coupling of environmental temperature fluctuations.

In the high-frequency range, the main noise mitigation has been achieved with the reduction of the Phasemeter additive noise, enhancing the signal amplitudes within the ADC range, and by matching the peak-to-peak values across the channels (section \ref{sec:pm_hfnoise}). The noise budget is dominated by the readout electronics, with a smaller contribution of the detector noise. Figures \ref{fig:nonrec_DFBL_FBBL}-\ref{fig:nonrec_FBBL_FSFBL} show that the measured noise floors are higher compared to the respective noise budgets. This discrepancy is possibly due to an underestimation during the calculation of the latter. We assumed all noise sources to be uncorrelated. However, they might be partially correlated, which would result in an increase of the noise floor level. We achieved with $\mathrm{\Phi_{DFBL}}- \mathrm{\Phi_{FBBL}}$ the best performance with a noise floor of 0.57 \text{pm}/$\sqrt{\text{Hz}}$ ($\mathrm{3.35 \  \mathrm{\upmu}rad /\sqrt{\text{Hz}}}$). 

In the center-frequency range, the performance is reached with active power and frequency stabilization (section \ref{sec:LaserNoise}). Additionally, for the DFBL and FBBL implementations, balanced detection is required to reduce the shoulder-shaped noise caused by the setup parasitic straylight (section \ref{sec:straylight}). The noise budget is dominated by the readout electronics, which is larger in the combinations including the FSFBL implementation since it operates with frequencies close to the bandwidth limitations (section \ref{sec:readoutNoise_LF}). Nonetheless, the measured noise lies within the noise budget, with $\mathrm{\Phi_{DFBL}}- \mathrm{\Phi_{FBBL}}$ and $\mathrm{\Phi_{FSFBL}}- \mathrm{\Phi_{DFBL}}$ only exceeding marginally the requirement at frequencies below 10 mHz.

In the low-frequency range, the noise has effectively decreased with the reduction of the environmental temperature fluctuations and a low drift (section \ref{sec:EnvTemp}), and with the mitigation of crosstalk between the TIA channels (section \ref{sec:TIA_HF}) and the photodetectors (section \ref{sec:detectorCrosstalk}). We observe a residual coupling, presumably from the environmental temperature fluctuations into our measurements, illustrated by the peak at 0.6 mHz. As previously explained, this matches with the air-conditioning cycle of the HVAC, and a potential coupling point would be in the Phasemeter. To further validate the thermal origin of this coupling, figure \ref{fig:nonrec_performance} includes a segment of the measurement $\mathrm{\varphi_{NR,FBBL} - {\varphi_{NR,FSFBL}}}$ during a period where the temperature stability, $\tilde{T}$, was much better (dashed green). During this time segment, the HVAC air-conditioning did not exhibit the default cycling behavior, corroborated by the absence of this oscillation in the time series of both the laboratory temperature sensors and the HVAC temperature monitor. As a result, the measurement does not feature the peak at 0.6 mHz. However, although this has been observed in numerous occasions, this result is not fully representative of the current performance since these are not the nominal laboratory conditions. 

With regards to the Backlink noise, we distinguish between the fiber-based and the free-beam implementations. For the fiber-based, we refer only to the DFBL, as the FSFBL is designed to be insensitive to the fiber backscatter coupling. The LFN that is coupled via this effect, which contributes on the noise in the center-frequency range, is mitigated with an active frequency stabilization (section \ref{sec:LFNcoupling}). In addition, the temperature-to-phase coupling via fiber backscatter, which contributes in the low-frequency range, is reduced with balanced detection (section \ref{sec:LFNcoupling}). For the free-beam implementation, an optimal design and integration of the DWS control loop (section \ref{sec:DWS}) is necessary to bring the performance to the required sensitivity levels. However, the results in this study are obtained without active rotation of the optical benches. This motion can potentially degrade the performance due to a reduction of the interferometric contrast if the alignment is not optimally controlled. The rotation of the benches to study the FBBL under full-operational conditions is planned for the near term. 

Finally, it is worth noting that our measurements combine the non-reciprocity of two Backlinks, suggesting that our results can be considered an upper limit only under the assumption of uncorrelated noise for the two links. At the same time, because of being a combined noise, the non-reciprocity of at least one --and potentially both-- Backlink is presumably better than the shown differential performance.

\section{Relevance for the LISA mission}
\label{sec:relevance}
\smallskip

In this section, we evaluate the DFBL implementation relative to the current Backlink baseline in LISA, referred here as LISA-PRDS. We present our conclusion that the DFBL can be considered as an upper limit for the nominal performance of the LISA-PRDS. First, we outline the differences on the fiber backscatter coupling into the main beam, as this is considered to be the dominant Backlink noise source on the Test Mass (TM) and Reference (REF) interferometers in LISA. Then, we briefly discuss technical aspects of the testbed implementation that may degrade the performance of the DFBL compared with the baseline configuration in LISA. Finally, we conclude by highlighting the main contributions of this study to the LISA mission.

\subsection{DFBL as an upper limit for the nominal performance of the LISA-PRDS}

Given the importance of the backscatter contribution to the local interferometers in LISA, it is essential to identify and analyze the key parameters in the 3BL that are likely to have a major impact on this effect. First, we discuss the contribution of the Backlink fiber and FIOS. The DFBL uses the Nufern PM980-XP, with backscatter values of 0.23\,ppm/m \cite{Xu:24} and fiber length of 2.6 m, which results in a total contribution of 0.6\,ppm. However, our analysis of section \ref{sec:backlinkNoise} suggests a backscatter contribution in the order of 7\,ppm. This higher value is not unlikely, and our current hypothesis assumes that it is originated from the FIOS. In the 3BL, we use a FIOS that was designed in-house. This differs from the FIOS used in LISA, designed at the University of Glasgow, due to the presence of an air gap between the lens and the main body of the FIOS, which potentially introduces more backscatter than the budgeted 1-2 ppm for LISA \cite{RohrPhD}. Furthermore, different power splitting ratios in the setup lead to a higher backscatter contribution in the 3BL interferometers compared to LISA. The measured backscatter-to-signal (b/s) power ratio in the 3BL is $\eta_{\text{BS}}^{\text{3BL}}=8.75\cdot10^{-7}$ (see section \ref{sec:backlinkNoise}), which is higher than the expected in the TM interferometer in LISA. In addition to this, we need to also consider the features in the optical testbed that differ between the 3BL and LISA. With respect to the optical bench, the main differences that can worsen the performance of the 3BL are the optics material and the bonding technique. The 3BL uses commercial optics and UV-glue bonding technique, while LISA uses space-qualified optics with better coating and hydroxide catalysis bonding technique, presumably resulting in a better stability of the optical bench. Moreover, the 3BL incorporates 95:5 attenuators and Faraday isolators for the incoming beams, L1 and L2, to prevent the light transmitted through the Backlink from traveling back to the input couplers, which would create backscatter in the input fibers that would enter the experiment, as this was identified as the primary limitation in the previous fiber-based Backlink experiment \cite{paperFleddermann2018}. However, the incorporation of the Faraday isolator can influence the impact of backscatter if the alignment deteriorates over time. 
Nevertheless, this is not representative for LISA, as it will not incorporate the Faraday isolator, but instead use a higher attenuator of \textgreater{99:1} from the telescope pick-off. This will result in a higher optical power reaching the input fibers. 

The results shown in figure \ref{fig:nonrec_DFBL_FBBL} suggest that, without balanced detection, the performance is limited likely by scatter contributions. Our current understanding indicates that the shoulder-shaped noise in the center-frequency range is expected to not be directly related to the Backlink fiber, requiring further investigation (see section \ref{sec:straylight}). The dominant contribution in the low-frequency part is likely originated from outcoupling FIOS backscatter, experiencing twice the fiber dynamics (see section \ref{sec:backlinkNoise}), with a power ratio higher than the budgeted value for LISA, as mentioned in the previous paragraph. 
 
As a conclusion, even though the 3BL is operated under better in-band temperature stability (see section \ref{sec:EnvTemp}) and it uses a shorter fiber, if we consider the higher backscatter contribution at the interferometer and the increased temperature-to-phase coupling (see section \ref{sec:T2Pscatter}, the performance can be considered an upper limit for LISA without balanced detection. Despite this, the results in figure \ref{fig:nonrec_DFBL_FBBL} also show that the DFBL reaches sub-picometer non-reciprocity levels over most of the measurement band when balanced detection is applied in post-processing.

\subsection{Main contributions for LISA}

The results obtained with the 3BL have contributed in evaluating the feasibility of a straightforward fiber-based Backlink implementation in LISA, confirming the relevance of the 3BL as a testbed for the LISA-PRDS. The measurements obtained with the 3BL identified temperature drifts to induce an additional non-linear coupling of the backscatter phase, as explained in section \ref{sec:T2Pscatter}.  
This observation contributed to establish a temperature drift requirement for the LISA Backlink. Additionally, we observe a noise suppression with balanced detection, corresponding to an efficiency of about 97\%. This further validates the effectiveness of balanced detection in a direct fiber Backlink.

\section{Conclusion} 
\smallskip

We have experimentally demonstrated that the Three-Backlink Experiment is a high-performance testbed for studying three different Backlink designs --two fiber-based, DFBL and FSFBL, and one transmitted in free-space, FBBL-- in the context of the LISA mission. Through an extensive commissioning process, which involved the characterization of the contributing noise sources and their corresponding mitigation strategies, we achieved a low-noise performance in the experimental setup. We have provided the total noise budget for the implementation of each Backlink design, showing that the 3BL is capable of measuring non-reciprocity levels below $3 \ \text{pm}/\sqrt{\text{Hz}} \cdot ~u_\text{}(f)$ across the entire measurement band, and below the $1 \ \text{pm}/\sqrt{\text{Hz}} \cdot ~u_\text{}(f)$ requirement across most of it. The sensitivity is exceeded between 0.5 mHz$-$20 mHz in the worst case, and between 2 mHz$-$4 mHz in the best case. We attribute this to the insufficient bandwidth in the readout electronics, and a residual coupling of the environmental temperature fluctuations.

Additionally, we report the preliminary findings of the noise inherent to each Backlink design. We studied the effects of LFN, mitigated with an active frequency stabilization. For the fiber-based, we studied the effects of temperature induced fiber-length changes via fiber backscatter, mitigated with balanced detection, and the effect of temperature drifts in the backscattered phase. For the free-beam, we demonstrated that the current infrastructure, which involves an active control of the steering mirrors based on the DWS technique, is capable of measuring sub-picometer non-reciprocity levels under static conditions. This establishes a solid baseline for studying a free-beam scheme under a more realistic scenario with rotating optical benches, which is planned for the near term. 

Furthermore, we measured the differential non-reciprocity for all three Backlink combinations. The results agree with the corresponding noise budgets --with a minor discrepancy at higher frequencies, considered uncritical as the final performance remains below the requirement--, and with our understandings of the Backlink noise. We observed that the DFBL is limited by backscatter noise from the Backlink fiber, contrary to the FSFBL, which is designed to be insensitive to it, and that the FBBL reaches the requirement under static conditions. 

Finally, we have compared the DFBL implementation with the current Backlink baseline in LISA. Even though the parameters are not identical, the results give a strong experimental indication that a direct fiber Backlink with balanced detection should be able to reach the required performance in LISA.

\ack{\noindent The authors thank M Rohr for the development of the laser power stabilization and C Bode for enhancing our understanding of phasemeters. The authors are grateful to all the personnel of the Institute's Mechanical and Electronics Workshops for their assistance. J J Ho-Zhang further acknowledges A Weidner for fruitful discussions on electronics. The authors thank A Taylor and D Robertson of University of Glasgow for their constructive feedback. The authors gratefully acknowledge support by the European Space Agency (ESA) within the project "Phase Reference Distribution System" (8586/16/NL/BW) and the Deutsches Zentrum für Luft- und Raumfahrt (DLR) with funding from the Bundesministerium für Wirtschaft und Technologie (Project Ref. Number 50 OQ 12301, based on work done under Project Ref. Number 50 OQ 1801, 50 OQ 1301 and 50 OQ 0601).}
 
\roles{\noindent Conceptualization: Gerhard Heinzel, Katharina-Sophie Isleif, Oliver Gerberding, Karsten Danzmann \\
\noindent Formal analysis: Melanie Ast, Jiang Ji Ho-Zhang \\
\noindent Funding acquisition: Jens Reiche \\
\noindent Investigation: Jiang Ji Ho-Zhang, Melanie Ast, Lea Bischof, Michael Born, Daniel Jestrabek, Stefan Ast  \\
\noindent Software: Oliver Gerberding, Thomas S. Schwarze \\
\noindent Supervision: Gerhard Heinzel, Melanie Ast \\
\noindent Visualization: Jiang Ji Ho-Zhang, Melanie Ast\\
\noindent Writing – original draft: Jiang Ji Ho-Zhang, Melanie Ast (Section 4.1)\\
\noindent Writing – review  \&  editing: Jiang Ji Ho-Zhang, Melanie Ast, Lea Bischof\\
\noindent All authors read and approved the final manuscript.}

\data{\noindent Data generated or analyzed during this study are available from the corresponding author upon reasonable request.}


\providecommand{\newblock}{}

\end{document}